\documentclass{aastex62}
\usepackage{lineno}
\usepackage{float}
\usepackage{amssymb,amsmath,natbib,gensymb}
\usepackage{color}
\usepackage{ulem}
\usepackage{latexsym}
\usepackage{wasysym} 
\usepackage{booktabs}
\shorttitle{CNIa0.02}
\shortauthors{Chen et al.}
\begin{document}

\title{The First Data Release of CNIa0.02 -- A Complete Nearby (Redshift $<0.02$) Sample of Type Ia Supernova Light Curves\footnote{This paper includes data gathered with the 6.5 meter Magellan Telescopes located at Las Campanas Observatory, Chile.}}
\author[0000-0003-0853-6427]{Ping Chen}\affil{Kavli Institute for Astronomy and Astrophysics, Peking University, Yi He Yuan Road 5, Hai Dian District, Beijing 100871, China}\affil{Department of Astronomy, School of Physics, Peking University, Yi He Yuan Road 5, Hai Dian District, Beijing 100871, China}
\author[0000-0002-1027-0990]{Subo Dong}\affil{Kavli Institute for Astronomy and Astrophysics, Peking University, Yi He Yuan Road 5, Hai Dian District, Beijing 100871, China}
\author{C.~S. Kochanek}
\affil{Department of Astronomy, The Ohio State University, 140 West 18th Avenue, Columbus, OH 43210, USA}
\affil{Center for Cosmology and Astroparticle Physics, The Ohio State University, 191 W. Woodruff Avenue, Columbus, OH 43210, USA}
\author{K.~Z. Stanek}
\affil{Department of Astronomy, The Ohio State University, 140 West 18th Avenue, Columbus, OH 43210, USA} 
\affil{Center for Cosmology and Astroparticle Physics, The Ohio State University, 191 W. Woodruff Avenue, Columbus, OH 43210, USA}
\author{R.~S. Post}\affil{Post Observatory, Lexington, MA 02421, USA}
\author{M.~D. Stritzinger}\affil{Department of Physics and Astronomy, Aarhus University, Ny Munkegade 120, DK-8000 Aarhus C, Denmark}
\author{J. L. Prieto}
\affil{N\'ucleo de Astronom\'ia de la Facultad de Ingenier\'ia y Ciencias, Universidad Diego Portales, Av. E\'jercito 441, Santiago, Chile}\affil{Millennium Institute of Astrophysics, Santiago, Chile}
\author[0000-0003-3460-0103]{Alexei V. Filippenko}\affil{Department of Astronomy, University of California, Berkeley, CA 94720-3411, USA}\affil{Miller Senior Fellow, Miller Institute for Basic Research in Science, University of California, Berkeley, CA 94720, USA}
\author{Juna A.~Kollmeier}\affil{Observatories of the Carnegie Institution for Science, 813 Santa Barbara Street, Pasadena, CA 91101, USA}
\author{N.~Elias-Rosa}\affil{Institute of Space Sciences (ICE, CSIC), Campus UAB, Carrer de Can Magrans s/n, 08193 Barcelona, Spain}\affil{Institut d'Estudis Espacials de Catalunya (IEEC), c/Gran Capit\'a 2-4, Edif. Nexus 201, 08034 Barcelona, Spain}
\author{Boaz Katz}\affil{Department of Particle Physics and Astrophysics, Weizmann Institute of Science, Rehovot 76100, Israel}
\author{Lina Tomasella}\affil{INAF - Osservatorio Astronomico di Padova, Vicolo dell'Osservatorio 5, I-35122 Padova, Italy}
\author{S.~Bose}\affil{Department of Astronomy, The Ohio State University, 140 West 18th Avenue, Columbus, OH 43210, USA}\affil{Center for Cosmology and Astroparticle Physics, The Ohio State University, 191 W. Woodruff Avenue, Columbus, OH 43210, USA}
\author{Chris Ashall}\affiliation{Institute for Astronomy, University of Hawaii at Manoa, 2680 Woodlawn Drive, Honolulu, HI 96822,USA}
\author{S.~Benetti}\affil{INAF - Osservatorio Astronomico di Padova, Vicolo dell'Osservatorio 5, I-35122 Padova, Italy}
\author{D.~Bersier}\affil{Astrophysics Research Institute, Liverpool John Moores University, 146 Brownlow Hill, Liverpool L3 5RF, UK}
\author{Joseph Brimacombe}\affiliation{Coral Towers Observatory, Queensland, Australia}
\author{Thomas G.~Brink}\affil{Department of Astronomy, University of California, Berkeley, CA 94720-3411, USA}
\author{P.~Brown}\affil{George P. and Cynthia Woods Mitchell Institute for Fundamental Physics \& Astronomy, USA}
\author{David~A.~H.~Buckley}\affil{South African Astronomical Observatory, P.O. Box 9, Observatory 7935, Cape Town, South Africa}
\author{Enrico Cappellaro}\affil{INAF - Osservatorio Astronomico di Padova, Vicolo dell'Osservatorio 5, I-35122 Padova, Italy}
\author{Grant W. Christie}\affil{Auckland Observatory, Auckland, New Zealand}
\author{Morgan Fraser}\affil{School of Physics, O'Brien Centre for Science North, University College Dublin, Belfield, Dublin 4, Ireland}
\author[0000-0002-1650-1518]{Mariusz Gromadzki}\affil{AstronomicalÂ Observatory, University of Warsaw, Al.Â UjazdowskieÂ 4, 00-478 Warszawa, Poland}
\author[0000-0001-9206-3460]{Thomas~W.-S.~Holoien}\altaffiliation{NHFP Einstein Fellow} \affiliation{The Observatories of the Carnegie Institution for Science, 813 Santa Barbara St., Pasadena, CA 91101, USA}
\author{Shaoming Hu}\affil{Shandong Provincial Key Laboratory of Optical Astronomy and Solar-Terrestrial Environment, Institute of Space Sciences, Shandong University, Weihai 264209, China}
\author{Erkki Kankare}\affil{Tuorla observatory, Department of Physics and Astronomy, University of Turku, FI-20014 Turku, Finland}
\author{Robert Koff}\affil{Antelope Hills Observatory, 980 Antelope Drive West, Bennett, CO 80102, USA}
\author{P.~Lundqvist}\affil{The Oskar Klein Centre, Department of Astronomy, Stockholm University, AlbaNova, SE-10691, Stockholm, Sweden}
\author{S.~Mattila}\affil{Tuorla observatory, Department of Physics and Astronomy, University of Turku, FI-20014 Turku, Finland}
\author{P.~A.~Milne}\affil{Department of Astronomy/Steward Observatory, 933 North Cherry Avenue, Rm. N204, Tucson, AZ 85721-0065, USA}
\author{Nidia Morrell}\affil{Las Campanas Observatory, Carnegie Observatories, Casilla 601, La Serena, Chile}
\author{J. A. Mu\~noz}\affil{Departamento de Astronom\'{\i}a y Astrof\'{\i}sica, Universidad de Valencia, E-46100 Burjassot, Valencia, Spain}\affil{Observatorio Astron\'omico, Universidad de Valencia, E-46980 Paterna, Valencia, Spain}
\author{Robert Mutel}\affil{Department of Physics and Astronomy, University of Iowa, Iowa City, IA 52242}
\author{Tim Natusch}\affil{Institute for Radio Astronomy and Space Research (IRASR), AUT University, Auckland, New Zealand}
\author{Joel Nicolas}\affil{Io Variablles-CCD group, 364 chemin de Notre Dame,06220 Vallauris,France}
\author{A.~Pastorello}\affil{INAF - Osservatorio Astronomico di Padova, Vicolo dell'Osservatorio 5, I-35122 Padova, Italy}
\author{Simon Prentice}\affiliation{School of Physics, Trinity College Dublin, The University of Dublin, Dublin 2, Ireland}
\author{Tyler Roth}\affil{Department of Physics and Astronomy, University of Iowa, Iowa City, IA 52242}
\author{B.~J.~Shappee}\affil{Institute for Astronomy, University of Hawaii at Manoa, 2680 Woodlawn Drive, Honolulu, HI 96822,USA}
\author{Geoffrey Stone}\affil{CBA Sierras, 44325 Alder Heights Road, Auberry CA 93602 USA}
\author[0000-0003-2377-9574]{Todd A. Thompson}
\affil{Department of Astronomy, The Ohio State University, 140 West 18th Avenue, Columbus, OH 43210, USA} 
\affil{Center for Cosmology and Astroparticle Physics, The Ohio State University, 191 W. Woodruff Avenue, Columbus, OH 43210, USA}
\author[0000-0001-6213-8804]{Steven Villanueva}\altaffiliation{Pappalardo Fellow}\affil{Massachusetts Institute of Technology, Cambridge, MA 02139 USA}
\author{WeiKang Zheng}\affil{Department of Astronomy, University of California, Berkeley, CA 94720-3411, USA}

\correspondingauthor{Subo Dong}
\email{dongsubo@pku.edu.cn}

\begin{abstract}
The CNIa0.02 project aims to collect a complete, nearby sample of Type Ia supernovae (SNe~Ia) light curves, and the SNe are volume-limited with host-galaxy redshifts  $z_{\rm host}<0.02$. The main scientific goal is to infer the distributions of key properties (e.g., the luminosity function) of local SNe~Ia in a complete and unbiased fashion in order to study SN explosion physics. We spectroscopically classify any SN candidate detected by the All-Sky Automated Survey for Supernovae (ASAS-SN) that reaches peak brightness $<16.5$\,mag. Since ASAS-SN scans the full sky and does not target specific galaxies, our target selection is effectively unbiased by host-galaxy properties.  We perform multi-band photometric observations starting from the time of discovery. In the first data release (DR1), we present the optical light curves obtained for 247 SNe from our project (including 148 SNe in the complete sample), and we derive parameters such as the peak fluxes, $\Delta m_{15}$ and $s_{BV}$.
\end{abstract}


\section{Introduction}

The explosion mechanism and progenitors of Type Ia supernovae (SNe~Ia) are basic but open questions in astrophysics. There are several proposed channels, but no agreement as to which or even how many of the channels dominate \citep[see, e.g.,][]{Maoz2014}. SNe~Ia span about an order of magnitude in peak luminosities and in the masses of synthesized $^{56}$Ni that power the radiation. It is also debated whether this range in properties represents one continuous population or more than one overlapping but distinct populations. On the one hand, the main properties of SNe~Ia appear to be continuous across the whole luminosity range. \citet{Phillips1993} found that the peak luminosity of SNe~Ia is tightly correlated with the light-curve shape characterized by the $B$-band post-peak decline rate $\Delta m_{15}(B)$, and this width-luminosity relation (WLR) is the foundation for using SNe~Ia as cosmological distance indicators. Many properties of their light curves \citep[see, e.g.,][]{Phillips1993, Phillips2012, Burns2014, Bulla2020} and spectra \citep[see, e.g.,][]{Nugent1995, Branch2009} also appear to be continuous. On the other hand, the possible existence of more than one populations of SNe~Ia has been long discussed \citep[e.g.,][]{BranchMiller1993}, including recent claims of bimodality in the luminosity function \citep[via the proxy of $\Delta m_{15}(B)$; see, e.g.,][]{Ashall2016, Hakobyan2020}, existence of two classes of fast-declining SNe Ia \citep{Dhawan2017}, and distinct near-ultraviolet (NUV)-optical \citep{Milne2013} and early-time optical \citep{Stritzinger2018} colors.

There have been tremendous efforts to obtain high-quality multi-band light curves of large samples of nearby SNe~Ia \citep[e.g.,][]{Hamuy1996, Riess1999, Jha2006, Hicken2009, Ganeshalingam2010, Contreras2010, Stritzinger2011, Hicken2012, Krisciunas2017, Foley2018, Stahl2019}. However, collecting a complete and unbiased nearby sample has only been made possible recently, thanks to the advent of wide-field time-domain surveys that do not target specific galaxies, such as the All-Sky Automated Survey for SuperNovae (ASAS-SN; \citealt{Shappee2014, Kochanek2017}), the {\it Gaia} transient survey \citep{GaiaAlerts2021}, the Palomar Transient Factory \citep{Law2009} and its successor the Zwicky Transient Facility (ZTF; \citealt{Kulkarni2016,ZTFbright}), the Asteroid Terrestrial-impact Last Alert System (ATLAS; \citealt{Tonry2011, Tonry2018b}), the Mobile Astronomical System of TElescope Robots (MASTER; \citealt{Gorbovskoy2013}), OGLE Transients Detection System (OTDS; \citealt{Wyrzykowski2014}), the Pan-STARRS Survey for Transients (PSST; \citealt{Huber2015, Chambers2016}), and the Catalina Real-Time Transient Survey (CRTS; \citealt{Drake2009}). Compared to other un-targeted surveys, ASAS-SN is a dedicated survey with a main goal to search for bright, nearby supernovae by scanning the entire visible sky at approximately nightly cadence (a cadence 2-3 night down to $\sim 17$\,mag prior to the expansion in 2017 and nightly cadence down to $\sim 18.5$\,mag after the expansion). The {\it Gaia} transient survey has the limiting magnitude down to $20.7$\,mag, and it is also an all-sky transient survey, while it has a very uneven cadence across the sky, which can be up to months. Most other surveys do not have full-sky coverage, while many of them have access to a large fraction of the sky at typical cadence on the order of days with deeper limiting magnitudes (given in the parentheses following the survey names) compared with ASAS-SN: ZTF ($\sim20.5$\,mag), Pan-STARRS ($\sim21.8$\,mag), MASTER ($\sim 20$\,mag), ATLAS ($\sim 20$\,mag), CRTS ($\sim19.5$\,mag). For most un-targeted surveys, there is no attempt to make spectroscopic classifications for all detected candidates selected according to certain criteria to form a complete sample. Furthermore, many time-domain surveys are primarily carried out in single bands, so without additional systematic follow-up efforts, it is not possible to obtain color information that are critical for deriving host-galaxy extinction and constrain SN physics.

We carry out the CNIa0.02 project to collect a {\bf C}omplete, {\bf N}earby, and effectively unbiased sample of Type {\bf Ia} Supernovae at host-galaxy redshifts $z_{\rm host}<{\bf 0.02}$ with well-observed multi-band light curves. Our follow-up observations started in January 2015 and ended in January 2020, and the SNe observed between 2015 September 17 and 2019 January 31 followed the selection criteria of the complete sample discussed below. The main goal for constructing a complete sample that is unbiased toward host-galaxy properties is to enable reliable statistical inference on the distributions of photometric properties of SNe~Ia (e.g., luminosity, color, light-curve shape, and derived physical parameters) in the local universe and also to study their dependence on host-galaxy properties. 

To our knowledge, collecting and studying a complete sample in astronomy can be traced back to  \citet{Schmidt1968}, which studied a complete sample of quasars defined with an observed flux density limit to derive their spatial distribution and luminosity function. Since then, complete samples have been widely applied in many areas of astronomy, and for instance, the LOSS survey produced one of the most influential complete samples of SNe from targeted searches \citep{Leaman2011, Li2011a, Li2011b}. Such a complete sample is defined to include all objects that meet a certain set of well-defined selection criteria on observables, making it possible to derive quantitative completeness corrections for inferring the statistical distribution of intrinsic properties such as the luminosity function. For the complete sample of CNIa0.02, we adopt the following observational selection criteria: (a) host-galaxy redshifts $z<0.02$, (b) peak brightness $V_{\rm peak} <16.5$\,mag, and (c) detection by the ASAS-SN survey, that is, we not only include SNe discovered by ASAS-SN, but also SNe discovered first by others and later detected by ASAS-SN.  
 The ASAS-SN detections are nearly $100\%$ complete for SNe with peak brightness $<16.5$\,mag (see Appendix \ref{sec:completeness}), and the ASAS-SN sample also has minimal bias in host-galaxy properties or SN locations inside the hosts \citep{Holoien2017a, Holoien2017b, Holoien2017c, Holoien2019}. All of the SNe in DR1 have been spectroscopically classified by ASAS-SN or other groups. The complete sample includes all spectroscopic sub-classes that are known to follow the WLR of the SNe Ia population. These include Ia-91bg and Ia-91T sub-types but exclude SNe Iax and other peculiar SNe Ia-like objects that deviate from the WLR of SNe Ia (see Appendix \ref{sec:purity} for a detailed discussion). The redshifts derived from SN classification spectra generally have too large uncertainties for our purpose, so we adopt host-galaxy spectroscopic redshifts for our complete sample selection. Where host-galaxy redshifts were unavailable in the NASA/IPAC Extragalactic Database\footnote{\dataset[10.26132/NED1]{https://ned.ipac.caltech.edu}} (NED), we have also measured the host-galaxy redshifts directly to determine whether the SNe Ia belong to the complete sample. We do not exclude supernova candidates without apparent hosts from our selection (i.e., the ``hostless'' SN). In our project, ASASSN-18nt is the only hostless SN, which is an intra-cluster SN Ia located in the galaxy cluster Abell 0194 ($z=0.018$), and its peak brightness ($16.66\pm 0.02$) does not qualify our selection criterion for the complete sample. All of the SNe were followed photometrically, mainly in the optical bands (primarily $BVri$), but with near-infrared (IR) and {\it Swift} NUV observations of some objects as well. In this first data release (DR1) of CNIa0.02, we present optical light curves for 247 SNe~Ia observed between 2015 and 2020.  CNIa0.02 DR1 includes some SNe Ia which are not in the complete sample, and the complete sample has 148 SNe in total. We describe the overall project and the sample in Section~\ref{sec:program}, the data processing in Section~\ref{sec:data_reduction}, and the resulting light curves in Section~\ref{sec:result}. Our present results are summarized in Section~\ref{sec:summary}.

\section{Program Description and the Sample}
\label{sec:program}
We select our targets primarily based on ASAS-SN detections, and the complete sample was collected between  2015 September 17 and 2019 January 31.  We have also observed a few SNe~Ia before (since January 2015) and after this period (until January 2020), and they are included in DR1 but are not part of the complete sample. In the early phase of the complete sample collection, we attempted to observe all SNe~Ia with $z<0.034$ and a peak magnitude of $V_{\rm peak}<17$. Between October 2016 and January 2019, we restricted the complete sample to focus on SNe~Ia with $z<0.02$ and a peak magnitude of $V_{\rm peak}<16.5$ as shown in Fig.~\ref{fig:protocol} and discussed in Appendix~\ref{protocol}. The detection efficiency of the ASAS-SN survey has been evolving mainly owing to upgrades in hardware, and since 2015, the detection efficiency has been almost $100\%$ complete to $<16.5$\,mag (see Appendix~\ref{sec:completeness} for a detailed discussion on sample completeness). 

\begin{figure}[h]
\centering
\includegraphics[width=12cm]{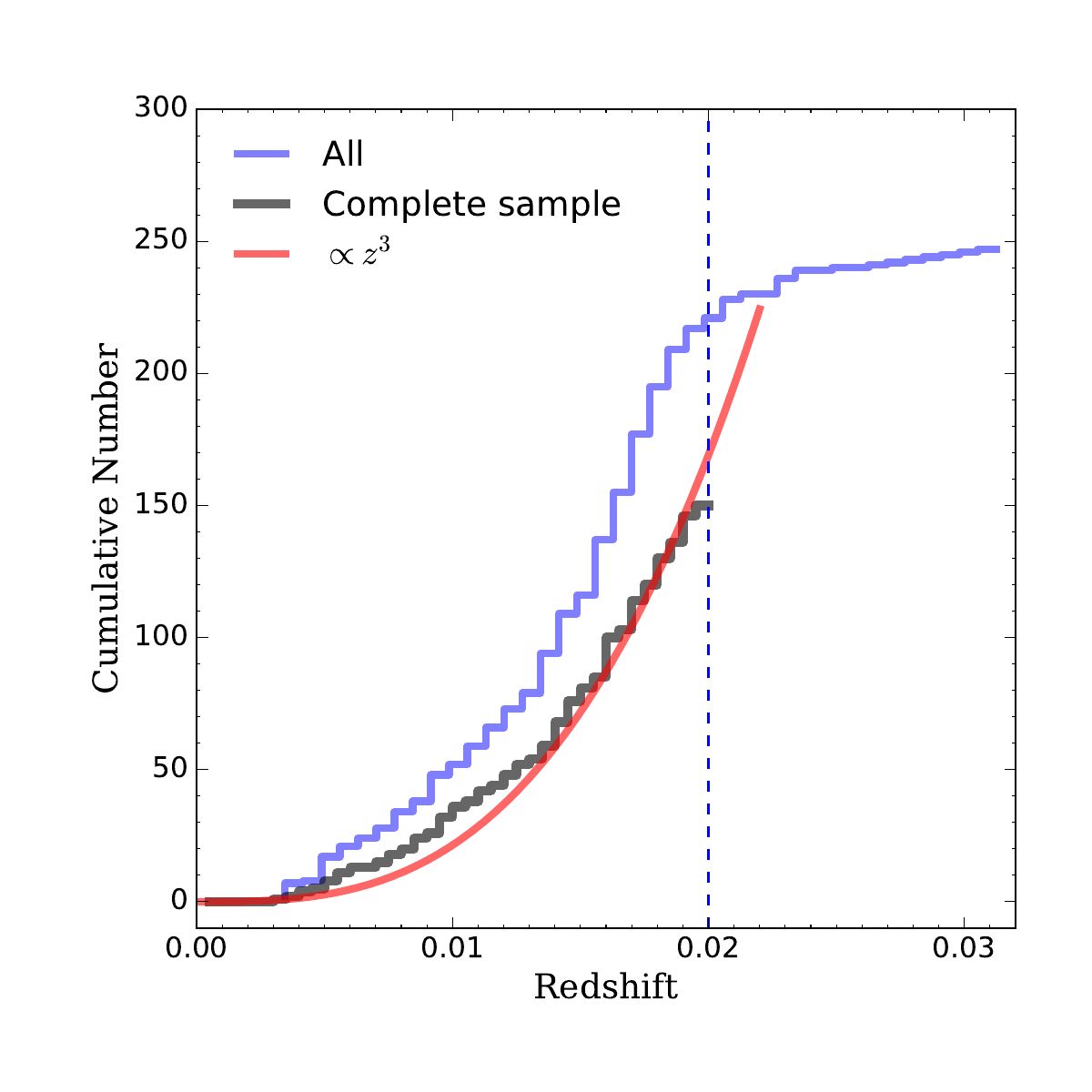}
\caption{{ The cumulative redshift distribution of all SNe~Ia (blue histogram) in DR1 and those in the complete sample (black histogram) from the CNIa0.02 project. The redshift limit of $z=0.02$ for the complete sample is indicated with a blue dashed vertical line. An illustrative $N \propto z^3$ is plotted as red line to indicate a simplified expectation from a volume-limited sample covering the full luminosity range by assuming a linear relation between distance and redshift. The distribution approximately follows the expectation for a volume-limited sample at $z\gtrsim0.01$, for which peculiar velocities are negligible compared to the Hubble expansion velocity. The apparent excess of SNe with $0.005\lesssim z \lesssim 0.013$ with respect to the volume-limited expectation is probably contributed by the effects of peculiar velocities at low redshift and/or fluctuations due to small-number statistics.}}
\label{fig:zdist}
\end{figure}

In Table~\ref{tab:sample}, we give the general information (names given by the survey groups, IAU names, equatorial coordinates, discovery dates, host-galaxy names, and heliocentric host redshifts) for all objects in the CNIa0.02 DR1, which includes objects that have follow-up data (regardless of whether they belong to the complete sample) or have been considered for follow-up observations (regardless of whether such data are obtained). The host-galaxy redshifts are from either NED or new measurements presented in Table~\ref{tab:redshift}. There are four SNe whose host-galaxy spectroscopic redshifts are not yet available, and for them, the redshifts determined from the supernova spectra are given in Table~\ref{tab:sample} and indicated with asterisks. Note that for all those four SNe, their peak magnitudes are fainter than 16.5, so they do not belong to the complete sample. We also provide additional information of $V$-band peak magnitudes (see Section~\ref{sec:lcparam} for how they are measured) and whether they were detected by ASAS-SN in Table~\ref{tab:sample}. The complete sample includes 148 SNe. In Figure~\ref{fig:zdist}, the cumulative distribution of host-galaxy redshifts of all SNe and those in the complete sample are shown as blue and black histograms, respectively, and the latter roughly follows the expectation for a volume-limited complete sample (shown as a red line) when the peculiar velocity is negligible compared to the Hubble expansion velocity (at $z\gtrsim0.01$). Note that our complete sample includes all SNe Ia selected by the observational criteria of $V_{\rm peak}<16.5$ and $z=0.02$, it does not include all SNe Ia at the dim end of the luminosity function ($\gtrsim -18.2$) near $z=0.02$, therefore it is not expected to exactly follow the distribution of a volume-limited complete sample covering the full luminosity range.

CNIa0.02 DR1 includes $V$-band and $g$-band photometry from the 14\,cm telescopes used to conduct the ASAS-SN survey.  Immediately after the discovery of a SN candidate that met our magnitude criteria, we started multi-band photometric observations, regardless of whether a spectroscopic classification was available then. For most objects, this data release contains follow-up photometry ending around 40--60\,days after the optical peak. For objects with bright galaxy backgrounds that require image subtractions, we took template images at least 300\,days after $B$-band peak, when the SN is typically more than $\gtrsim 7$\,mag below the peak. We have performed photometric follow-up observations using a number of telescopes ranging from $\sim0.3$\,m to $\sim2$\,m. In this data release, most data are in $BVri$ bands observed by 1-m telescopes of the Las Cumbres Observatory Global Telescope network (LCOGT; \citealt{Brown2013}) distributed over four sites covering both hemispheres,  two 0.6\,m telescopes in Sierra Remote Observatories (CA, USA) and Mayhill (NM, USA) of the Post Observatory (PO), and the 1.3\,m telescope of Small \& Moderate Aperture Research Telescope System (SMARTS; \citealt{Subasavage2010}). For supernovae found between October 2016 and March 2018, we carried out a follow-up program using the Ultra-Violet/Optical Telescope (UVOT; \citealt{Roming2005}) on {\it the Neil Gehrels Swift Observatory} ({\it Swift} \citealt{Gehrels2004}), and the UVOT $bv$-band data from that program are included in DR1. We also include some photometric data obtained from the 2\,m Liverpool Telescope (LT), 0.5\,m DEdicated MONitor of EXotransits and Transients (DEMONEXT; \citealt{Villanueva2018}), the 1-m telescope at WeiHai observatory of Shandong University (WHO; \citealt{Hu2014}), a 0.41-m telescope at A77 observatory, the Ohio State Multi-Object Spectrograph (OSMOS) on the 2.4\,m Hiltner Telescope at the MDM observatory, the Wide Field reimaging CCD (WFCCD) camera and direct imaging CCD camera SITe2K on the 2.5\,m du Pont telescope, and Alhambra Faint Object Spectrograph and Camera (ALFOSC) on the 2.56\,m Nordic Optical Telescope (NOT). The instrument specifications for the above facilities are described in Appendix~\ref{sec:instrument}. We plan to make other follow-up data collected by our project available in the future. 

\section{Data Processing}
\label{sec:data_reduction}

This data release contains the results of processing over 20,000 images from ground-based observations and also {\it Swift}-UVOT images. For ground-based data, we developed a photometric pipeline {\it PmPyeasy} to automatically process the images and obtain the photometry. The pipeline utilizes several external software packages that are all wrapped in a Python interface. The pipeline runs automatically by default, while it allows manual operations at any point when necessary. The pipeline uses {\it pyds9}\footnote{\url{http://hea-www.harvard.edu/RD/pyds9/}} to facilitate human inspections through XPA messaging to SAOImageDS9\footnote{\url{https://sites.google.com/cfa.harvard.edu/saoimageds9}}. It takes images that have already been pre-processed including bias removal and flat fielding. Below we outline our procedures, and at the end of the the section, we summarize our reduction of the UVOT data.

\subsection{Image Registration and Source Detection}
\label{sec:1ststep}
The pipeline distributes all the images to object-specific folders and adds information such as the filter, exposure time, and epoch to a database. Next, it removes cosmic rays using an implementation of the L.A.Cosmic algorithm \citep{Dokkum2001}, measures the full width at half maximum (FWHM) of the stellar profiles, and estimates the background value for each image. It then employs {\it PyRAF daofind} to generate a source catalog for each image.

\subsection{PSF Photometry and Image Subtraction}
\label{sec:2ndstep}
We perform point-spread-function (PSF) photometry for SNe that have negligible host-galaxy contaminations using DoPHOT \citep{Schechter1993, Alonso-Garca2012}. For each image, DoPHOT generates a PSF model automatically and yields magnitudes for point sources.

A large number of targets (102 out of 247 SNe) have significant host-galaxy background fluxes and require image subtraction. To perform image subtraction, the pipeline first matches point sources detected on the science image with those on the template image, and then the science image is astrometrically aligned to the same reference frame of the template image using the matched sources and resampled. The image subtraction is done with the High Order Transform of PSF ANd Template Subtraction package (HOTPANTS; \citealt{Becker2015}). The FWHMs of the template and resampled science image are used to determine the convolution direction: images with better seeings are convolved with the kernel for subtraction. We configured HOTPANTS to normalize the fluxes measured on all subtracted images to the template's flux scale. To perform photometry for targets after image subtraction, the pipeline first identifies isolated stars with high signal-to-noise ratios (SNRs) on the template image, and these stars are used to build a PSF model for each convolved image. Then PSF photometry is performed at the SN position on the subtracted image and for all the sources on the template image using the {\it PyRAF daophot} task.

In some cases, host-galaxy flux subtraction is required but image subtraction is not feasible when there are too few reference stars available in the observed field or template images are not available. If a SN is under such a circumstance and its host galaxy has a smooth profile that can be characterized by an isophote model (e.g., an elliptical galaxy), we devise a method to subtract the host-galaxy flux by incorporating an ellipse isophote modeling of the host galaxy. We adopt the following steps: (1) perform the usual PSF photometry with {\it PyRAF/daophot} for point sources (including the SN) within the region to be fitted by an isophote model; (2) { subtract the point sources from the image, and then} use the {\it isophote/ellipse} task from {\it PyRAF/stsdas} package to model the host-galaxy flux on the point-source-subtracted image; (3) subtract the best-fit isophote model from the original image and then perform PSF photometry for the stellar objects { on the galaxy-flux-subtracted image}; (4) steps (2) and (3) are then performed iteratively for three more times. {In each iteration the isophote model for the galaxy and the PSF photometry for the stellar objects are refined.} {This method has been used for the following targets with corresponding telescope/instruments given in the parentheses: 2017jfw (SMARTS), 2018ast (LCOGT 2m, PO, LT, MDM), ASASSN-18an (SMARTS), ASASSN-18en (SMARTS), 2016fnr (NOT), 2016gfr (NOT), 2016iuh (MDM), ASASSN-16la (MDM), ASASSN-17fr (du Pont/SITe2K).}

\subsection{Photometric Calibration}
For photometric calibration, we transform our photometry to the standard Johnson magnitudes ($BV$) in Vega system and SDSS magnitudes ($ri$) in AB magnitude system, respectively, using the reference stars with available calibrated magnitudes in the field. Since our targets cover the full sky, the preferred sources for reference stars should be an all-sky catalog with homogeneous photometric calibrations. We use the photometric system defined by the Pan-STARRS1 (PS1) survey \citep{Chambers2016}, which has a well-characterized photometric system, with transformations to other standard photometric systems available in \citet{Tonry2012}. The PS1 $3\pi$ Steradian Survey\citep{Chambers2016} has multi-band ($grizy_{P1}$) coverage of the sky with declinations $>-30\degree$, and we use photometry given in the Pan-STARRS1 DR1 MeanObject database \citep{Flewelling2020}. For the remaining quarter of the sky, we use the ATLAS All-Sky Stellar Reference Catalog (Refcat2), which was assembled from a variety of sources and brought onto the the same photometric system as Pan-STARRS1 \citep{Tonry2018}. Before being used for photometric calibrations of our targets, the PS1 (or Refcat2) magnitudes of the reference stars in the fields are first converted to Johnson $BV$ and SDSS $ri$ bands adopting the following transformations \citep{Tonry2012}:

\begin{displaymath}
B_{\rm Johnson} = g_{\rm PS1} + 0.212 + 0.556\, (g_{\rm PS1} - r_{\rm PS1}) + 0.034\, (g_{\rm PS1} - r_{\rm PS1})^2,
\end{displaymath}
\begin{displaymath}
V_{\rm Johnson} = r_{\rm PS1} + 0.005 + 0.462\, (g_{\rm PS1} - r_{\rm PS1}) + 0.013\, (g_{\rm PS1} - r_{\rm PS1})^2,
\end{displaymath}
\begin{displaymath}
r_{\rm SDSS} = r_{\rm PS1} - 0.001+ 0.004\, (g_{\rm PS1} - r_{\rm PS1}) +  0.007\,(g_{\rm PS1} - r_{\rm PS1})^2,
\end{displaymath}
\begin{displaymath}
i_{\rm SDSS} = i_{\rm PS1} - 0.005 + 0.011\,(g_{\rm PS1} - r_{\rm PS1}) + 0.010\,(g_{\rm PS1} - r_{\rm PS1})^2.
\end{displaymath}

In practice, we use reference stars brighter than 19\,mag in the field. For a target using PSF photometry, our measured magnitudes of the references are matched to standard magnitudes to derive a zero-point offset for each image. For a target using image subtractions, the flux scale of the template is calibrated using the references, and then all measured magnitudes are scaled to the same photometric system as the template. The photometric uncertainties are estimated by quadratically combining the photometric errors reported by DoPHOT or {\it PyRAF daophot} with those of the zero-point calibrations into the standard systems. The typical uncertainty of our calibrated photometry is $\sim 0.05$\,mag.

\subsection{{\it Swift} UVOT photometry}

In this section, we briefly describe how we perform {\it Swift} UVOT $bv$ photometry, and detailed discussions and results of our full {\it Swift} SNe Ia campaign will be given in a future paper. Processed {\it Swift} UVOT images are downloaded from the {\it Swift} Archive\footnote{https://heasarc.gsfc.nasa.gov/FTP/swift/}. We follow the same basic photometric procedures as described in \cite{Brown2014}. We use the calibration database (CALDB) version released on 2020-12-15, which includes the revised photometric zero points \citep{Breeveld2011} and latest time-dependent detector sensitivity. We follow the {\it Swift} UVOT standard photometric calibrations \citep{Poole2008,Breeveld2010} to extract the source counts on the science images and the host galaxy counts on the template images with an aperture with radius of $5\arcsec$. We subtract the host-galaxy contributions and then convert the source count rates to magnitudes in the UVOT-Vega system. 

\section{Results}
\label{sec:result}

\subsection{Light-Curve Data}
\label{sec:lcdata}

\begin{figure*}[h]
\includegraphics[width=17cm]{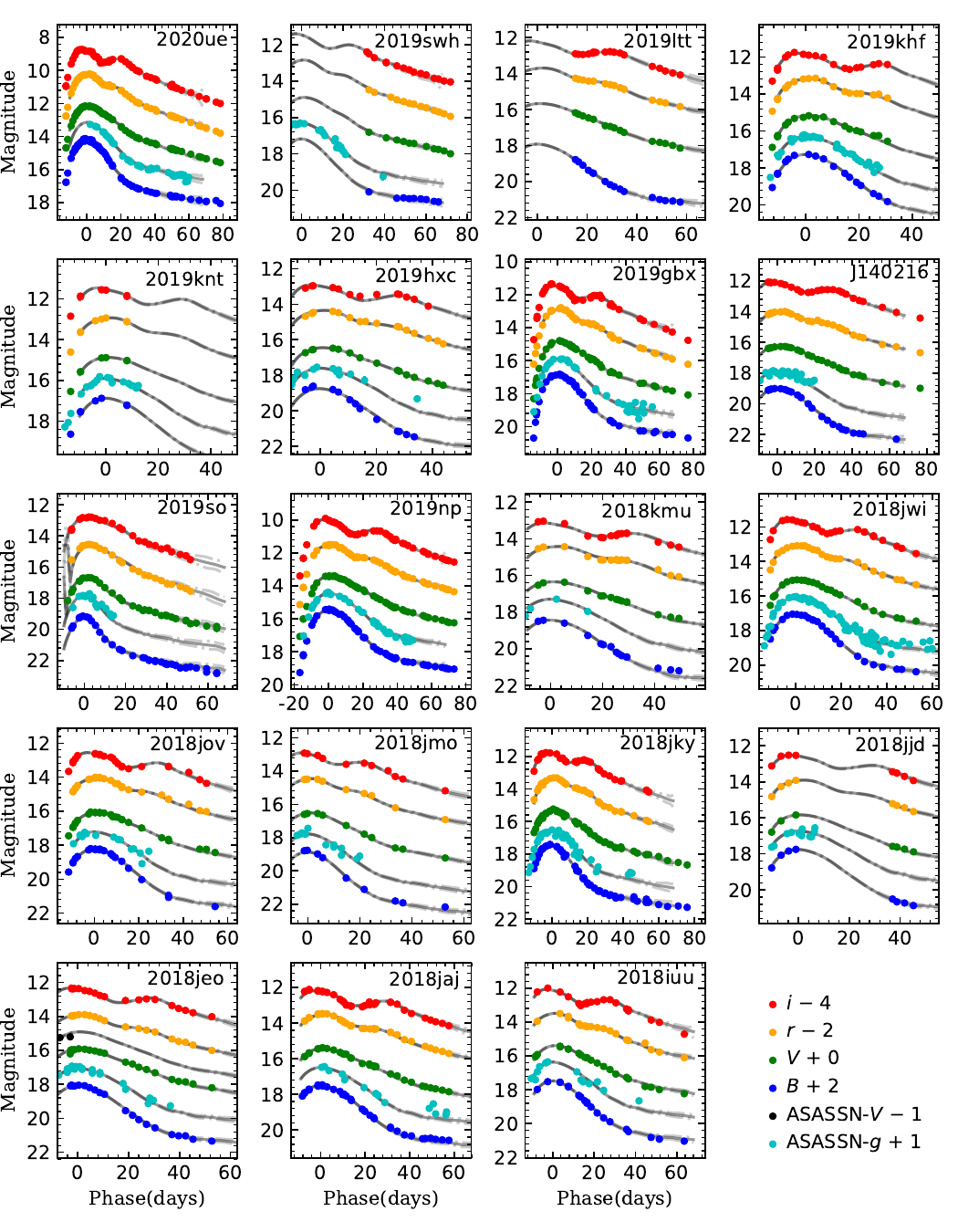}
\caption{Multi-band light curves of SNe~Ia in DR1. Here we show the light curves for 19 SNe~Ia in DR1 with the most recent discovery time. The black lines correspond to the fitting results using the {\it max\_model} fitting of SNooPy, and the corresponding best-fit parameters are given in Table~\ref{tab:lcparam_snpy_model} (see Section~\ref{sec:lcparam}). All phases in days are with respect to the time of $B$-band peak obtained from the {\it max\_model} fitting. A full version of this figure including all SNe~Ia with multi-bands light curves is available on the journal website.}
\label{fig:lcplot}
\end{figure*}

In this section, we present the optical light curves of 247 SNe~Ia. Most of them have ASAS-SN $Vg$-band light curves using image subtractions (see \citealt{Jayasinghe2018} for descriptions on the ASAS-SN image-subtraction photometry). For 219 SNe, we conducted $BVri$ follow-up observations with LCOGT 1-m and PO telescopes, and the light curves for all of them are included in DR1. $BV$ light curves for 24 SNe obtained with SMARTS 1.3\,m telescope are included in this data release. We also include $BVri$ light curves for several targets obtained from LCOGT 2\,m telescope, LT, DEMONEXT and A77 as well as relatively late-phase data for a small number of targets from Hiltner, du Pont, and NOT. The light curves are given in Table~\ref{tab:lightcurves}, which are the main result of CNIa0.02 DR1. In Figure~\ref{fig:lcplot}, we show the multi-band light curves up to 80\,days past $B$-band peak (or the time of discovery if peak time is not available). 

\subsection{Light-Curve Parameters}
\label{sec:lcparam}

As discussed in Section~\ref{sec:program}, $V$-band peak magnitude $V_{\rm peak} < 16.5$ is one of the criteria for the complete sample of CNIa0.02. To obtain the $V$-band peak magnitudes of SNe Ia presented in Table~\ref{tab:sample}, we used the SNooPy\footnote{\url{https://csp.obs.carnegiescience.edu/data/snpy/snpy}} \citep{Burns2011} software to fit (utilizing the {\it ``max\_model''}) the observed light curves with SNe Ia template light curves. The light curves are shifted in both phase and brightness to find the best match with a set of template light curves characterized by the color-stretch parameter $s_{BV}$, which is found to be tightly correlated with peak luminosity across the full range of SN~Ia decline rate \citep{Burns2014}. $s_{BV}$, $B$-band peak time $t_{\rm peak}(B)$ and the peak magnitudes in all bands involved are free parameters. {\it Swift} UVOT $bv$ data are not included in our fitting, except for two SNe (2017emq and 2017fbj) whose UVOT light curves have the essential coverage missed by other sites. Since the follow-up $V$-band data are generally more precise and have better coverage than ASAS-SN, we only include ASAS-SN $V$-band data in cases where follow-up $V$-band data are unavailable. During the fitting process, $>5\,\sigma$ outliers from the model were removed iteratively. The best-fit parameters ($t_{\rm peak}(B)$, $s_{BV}$, $B_{\rm peak}$, $g_{\rm peak}$, $V_{\rm peak}$, $r_{\rm peak}$, $i_{\rm peak}$) for 232 SNe Ia in CNIa0.02 are given in the {\it max\_model} section of Table~\ref{tab:lcparam_snpy_model}, and the corresponding best-fit models are displayed in Figure~\ref{fig:lcplot}. 

We also fit the data using SNooPy's {\it ``EBV\_model2''}, which can derive host-galaxy extinctions. The {\it EBV\_model2} method fit the light curves with the templates as described below 
\begin{equation} 
\begin{aligned} 
m_X(\phi) = T_X(\phi, s_{BV}) + M_X(s_{BV}) + \mu + K_{XY} 
+ R_X^{\rm MW} \cdot E(B-V)_{\rm MW} + R_X^{\rm host} \cdot E(B-V)_{\rm 
host} ,
\end{aligned} 
\end{equation} 
where $m_X$ is the observed magnitude in band $X$, $T_X(\phi, s_{BV})$ is the template light curve as a function of rest-frame phase $\phi$ and $s_{BV}$, $M_X(s_{BV})$ is the peak absolute magnitude of the SN with given $s_{BV}$, $\mu$ is the distance modulus in magnitudes, $K_{XY}$ is the cross-band k-correction from Y band to the observed X band, $E(B - V)_{\rm gal}$ and $E(B - V)_{\rm host}$ are galactic and host-galaxy color excess due to extinction, and $R_X^{\rm gal}$ and $R_X^{\rm host}$ are the ratios of total to selective extinction for the Milky Way and the host galaxy, respectively. Among the parameters listed above, $M_X(s_{BV})$, $K_{XY}$, $E(B-V)_{\rm MW}$, $R_X^{\rm MW}$, $R_X^{\rm host}$ are predetermined and provided by SNooPy, and $t_{\rm peak}(B)$, $s_{BV}$, $E(B-V)_{\rm host}$, and $\mu$ are free parameters in the fitting. $E(B-V)_{\rm MW}$ is obtained from the results of \cite{Schlafly2011} and the canonical $R_V^{MW}=3.1$ is adopted for the Milky Way. SNooPy has different sets of calibration results of the peak luminosity of SNe Ia, and we adopt $R_V^{\rm host}=1.729$ (corresponding to calibration$=5$ in SNooPy), which is the result of calibration by using SNe Ia covering the full range of $s_{BV}$ \citep{Burns2014}. Our dataset generally have the best coverage in $BVri$, and light curves in these bands are used in the {\it EBV\_model2} fitting for all objects, except for four objects (2018hkq, 2018htw, 2018kmu and 2019swh). When the $g$-band light curves provide coverage missed by other bands, they are also used in the fitting. We obtain the best-fit parameters ($t_{\rm peak}(B)$, $s_{BV}$, $E(B-V)_{\rm host}$, and $\mu$) for 212 SNe Ia, and they are listed in the {\it EBV\_model2} section of Table~\ref{tab:lcparam_snpy_model}. 

We also perform model-independent fitting to the well-covered SN~Ia light curves to directly derive parameters including the times and magnitudes of peak brightness and the decline rates in the $B$ and $V$ bands. The decline rate $\Delta m_{15}(X)$ \citep{Phillips1993} refers to the magnitude decline within 15\,days after peak brightness in a given filter $X$. We measure these parameters directly from the interpolated light curves in $B$ and $V$ band using a Gaussian process regression method, which has the advantage of allowing for the inclusion of uncertainty information and producing relatively unbiased estimates of interpolated values (see, e.g., \citealt{Lochner2016}). The results are given in Table~\ref{tab:lcparam}. Note that the fitting is performed without making host-galaxy extinction corrections, which may affect the derived $\Delta m_{15}$ for objects with high extinction \citep{Phillips1999}.

\begin{figure}
\centering
\includegraphics[width=12cm]{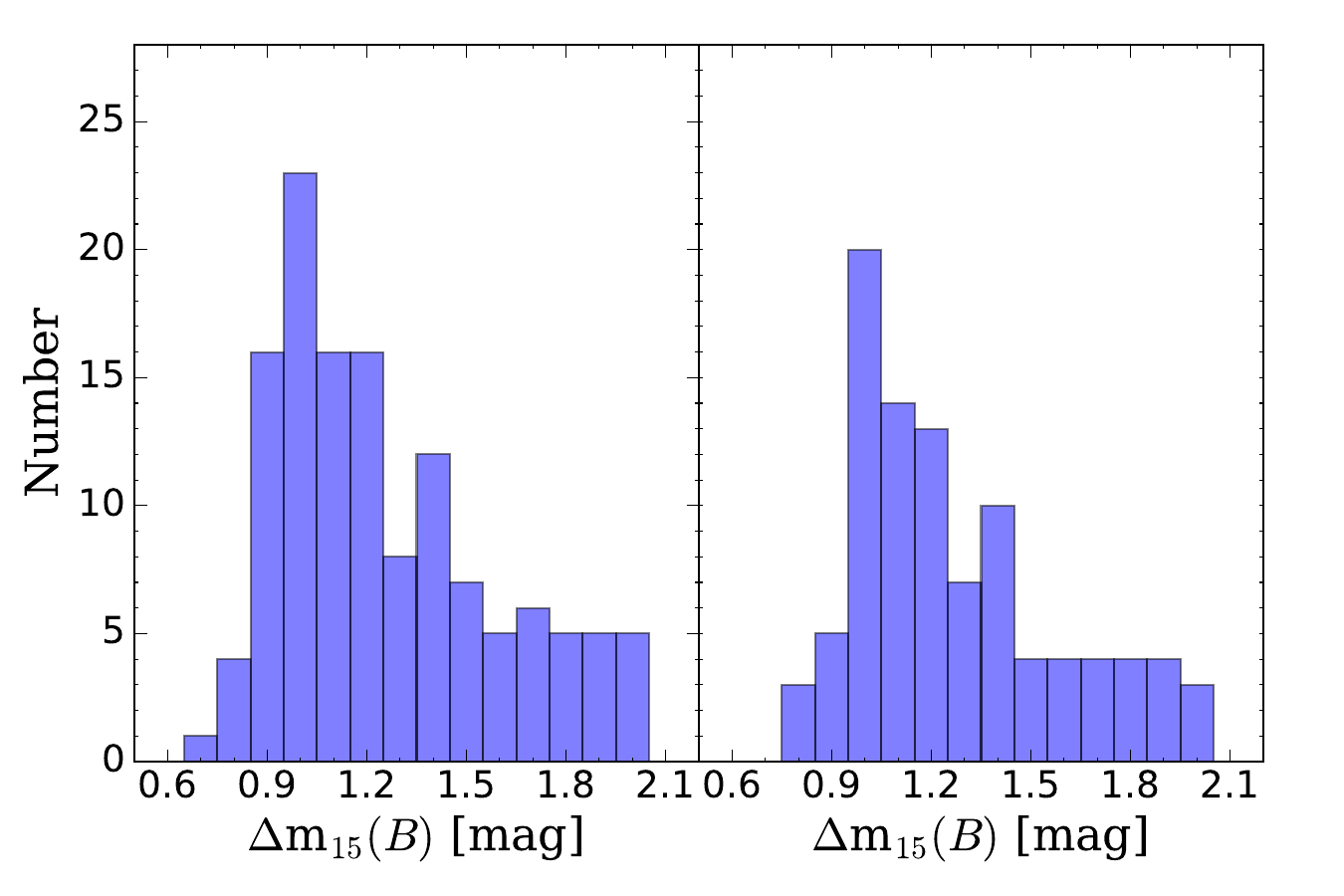}
\caption{Distribution of all available direct $\Delta m_{15} (B)$ measurements for 129 SNe Ia in DR1 (left panel) and 95 SNe Ia in the complete sample (right panel).}
\label{fig:m15Bdist}
\end{figure}

Figure~\ref{fig:m15Bdist} shows the histogram of all available direct $\Delta m_{15}(B)$ measurements for CNIa0.02 DR1. The left panel is for the SNe included in CNIa0.02 DR1, and the right panel is for those in the complete sample. Our objects include SNe Ia spanning the full range of $\Delta m_{15}(B)$ of the SN~Ia population from $\Delta m_{15}(B) \approx 0.7$ mag to $\Delta m_{15}(B) \approx 2.0$ mag. The complete sample consists of SNe Ia with $z<0.02$ and $V_{\rm peak}<16.5$\,mag, and due to the peak magnitude limits, our sample is not sensitive to all dim SNe Ia with high $\Delta m_{15}(B)$ values within the volume confined by $z<0.02$. Further work on quantifying detection efficiency within different $\Delta m_{15}(B)$ bins needs to be carried out to obtain the intrinsic distribution of $\Delta m_{15}(B)$ and other parameters for SNe Ia population.

\section{Summary}
\label{sec:summary}
CNIa0.02 aims to obtain a homogeneous and unbiased sample of nearby SNe~Ia with multi-band light curves to study the SNe~Ia population. In CNIa0.02 DR1, we present $247$ SNe with optical light curves, including {148} SNe in the complete sample. DR1 offers large and homogenous optical photometric data sets to systematically study the SNe~Ia population. In this paper, we present the first analysis of our dataset by extracting parameters such as $\Delta m_{15}(B)$. We plan to publish near-UV, near-IR, and late-phase photometric data in the future. Our multi-band light curves also allow us to derive host-galaxy extinction and luminosity, and in a forthcoming publication, we plan to make completeness correction and study the SN Ia luminosity function. CNIa0.02 provides a large and homogeneous dataset to infer the intrinsic distribution properties of SNe~Ia in the local universe to help answer basic questions regarding SN~Ia progenitor systems and explosion mechanisms. 

\acknowledgments

We thank Eric Peng and the anonymous reviewer for helpful suggestions. We acknowledge the Telescope Access Program (TAP) funded by the NAOC, CAS and the Special Fund for Astronomy from the Ministry of Finance. We acknowledge SUPA2019A (PI: M.D. Stritzinger) via OPTICON. CSK, KZS and BJS are supported by NSF grants AST-1515927, AST-1814440, and AST-1908570. M.D.S acknowledges funding from the Villum Fonden (project numbers 13261 and 28021). M.D.S is supported by a project grant (8021-00170B) from the Independent Research Fund Denmark.
 A.V.F.'s supernova group is grateful for financial assistance from the Christopher R. Redlich Fund, the TABASGO Foundation, and the Miller Institute for Basic Research in Science (U.C. Berkeley). A major upgrade of the Kast spectrograph on the Shane 3~m telescope at Lick Observatory was made possible through generous gifts from William and Marina Kast as well as the Heising-Simons Foundation. Research at Lick Observatory is partially supported by a generous gift from Google. We thank the staffs of the various observatories at which data were obtained for their excellent assistance. JLP is provided in part by FONDECYT through the grant 1191038 and by the Ministry of Economy, Development, and Tourism's Millennium Science Initiative through grant IC120009, awarded to The Millennium Institute of Astrophysics, MAS. MF acknowledges the support of a Royal Society - Science Foundation Ireland University Research Fellowship. BJS is also supported by NSF grants AST-1920392 and AST-1911074. MG is supported by the Polish NCN MAESTRO grant 2014/14/A/ST9/00121. Polish participation in SALT is funded by grant no. MNiSW DIR/WK/2016/07. SMH is supported by the Natural Science Foundation of Shandong province (No. JQ201702), and the Young Scholars Program of Shandong University (No. 20820162003).
Support for TW-SH was provided by NASA through the NASA Hubble Fellowship grant \#HST-HF2-51458.001-A awarded by the Space Telescope Science Institute, which is operated by the Association of Universities for Research in Astronomy, Inc., for NASA, under contract NAS5-26555. We thank the {\it Swift} PI Brad Cenko, the Observation Duty Scientists and the science planners for approving and executing our {\it Swift}/UVOT SNe Ia campaign.
 
\software{Astropy \citep{AstropyCollaboration2018}, PyRAF \citep{pyraf2012}, FITSH \citep{Pal2012}, ccdproc \citep{craig2017}}, HOTPANTS \citealt{Becker2015},  DoPHOT \citep{Schechter1993, Alonso-Garca2012}

\bibliographystyle{apj}
\bibliography{ms}

\begin{longrotatetable}
\begin{deluxetable}{llcccclccc}
\tabletypesize{\scriptsize}
\tablecolumns{9}
\tablewidth{0pt}
\tablecaption{General Properties of SNe Ia in CNIa0.02 DR1\label{tab:sample}}
\tablehead{
\colhead{Survey Name} & 
\colhead{IAU Name} &
\colhead{RA(J2000)}   & 
\colhead{Dec(J2000)}  &
\colhead{Discovery Date} &
\colhead{Host Galaxy} & 
\colhead{z$_{\rm host}$\tablenotemark{a}} &
\colhead{V$_{\rm peak}$\tablenotemark{b}} &
\colhead{ASAS-SN?\tablenotemark{c}}&
\colhead{Complete?\tablenotemark{d}}
}
\startdata
ASASSN-15aj &  & 10:52:53.261 & $-$32:55:34.86 & 2015-01-08 & NGC 3449 & 0.010921 & 14.77$\pm$0.02 & Y & N \\
ASASSN-15ak &  & 00:12:01.546 & +26:23:37.284 & 2015-01-09 & UGC 00110 & 0.015034 & 14.70$\pm$0.01 & Y & N \\
ASASSN-15db &  & 15:46:58.69 & +17:53:02.22 & 2015-02-15 & NGC 5996 & 0.010998 & 14.55$\pm$0.03 & Y & N \\
ASASSN-15eb &  & 08:06:07.399 & $-$22:33:48.852 & 2015-02-26 & ESO 561- G 012 & 0.016481 & 15.82$\pm$0.05 & Y & N \\
J073615$\dagger$ & 2015F & 07:36:15.76 & $-$69:30:23.0 & 2015-03-09 & NGC2442 & 0.00489 & 13.31$\pm$0.02 & Y & N \\
J150530$\dagger$ & 2015bp & 15:05:30.09 & +01:38:02.2 & 2015-03-16 & NGC 5839 & 0.004069 & 13.90$\pm$0.01 & Y & N \\
ASASSN-15ga &  & 12:59:27.293 & +14:10:15.78 & 2015-03-30 & NGC 4866 & 0.006631 & 15.08$\pm$0.03 & Y & N \\
ASASSN-15go &  & 06:11:30.401 & $-$16:29:04.596 & 2015-04-06 & WISEA J061130.50-162908.3 & 0.018923 & 16.04$\pm$0.09 & Y & N \\
\enddata
\end{deluxetable}
\tablenotetext{\dagger}{These names are used for brevity, and their corresponding full names are listed below.}
\tablenotetext{}{J020622: PSN J02062253-5201267, J122100: PSN J122100.9-533150.1, J015053: PSN J01505356-3600308, J103747: MASTER OT J103747.94-270507.2, J010720: PSN J01072038+3223598, J215050: PSN J21505094-7020289, J150530: PSN J15053007+0138024, J110533: PSN J110533.80+194118.7, J150915: CSS170619:150915-112003, J140216: PSN J140216.0-533228.8, J141551: MASTER OT J141551.21-480802.6, J213123: PSN J21312375+4336312, J033333: MASTER OT J033333.26-623314.7, J114925: PSN J11492548-0507138, J073615: PSN J07361576-6930230, J112345: PSN J11234588-0106212 .}
\tablenotetext{a}{Host-galaxy heliocentric spectroscopic redshifts taken from the NASA/IPAC Extragalactic Database (NED) or from new spectroscopic measurements in Table~\ref{tab:redshift}. If the host-galaxy spectroscopic redshift is not available, then the SN spectroscopic redshift is displayed here instead and indicated with asterisk. ASASSN-18nt (2018ctv) was discovered in the galaxy cluster Abell 0194 \citep{Chen2018ATel11762}, which was found to be not associated with any obvious galaxy in the cluster but located in the intra-cluster light appearing to bridge between the galaxy pair NGC545+547 and NGC541 \citep{Moral-Pombo2018ATel12313}. Here we adopt the redshift of the galaxy cluster for ASASSN-18nt \citep{Struble1999}.}
\tablenotetext{b}{Peak magnitudes in $V$ band obtained from template fitting with {\it max\_model} in SNooPy. For targets without successful template fitting results, if they are detected in ASAS-SN data and with redshift $z<0.02$, the upper limits for the peak magnitudes derived from available data are reported here. \cite{Dong2018} estimated $V_{\rm max}\sim 15.7$ for 2016brx by matching its data to the light curves of SN 1991bg.}
\tablenotetext{c}{Whether the SN was detected by the ASAS-SN survey.}
\tablenotetext{d}{Whether the SN belongs to our complete sample.}
\end{longrotatetable}

\begin{deluxetable}{llllc}
\tabletypesize{\footnotesize}
\tablecolumns{15}
\tablewidth{0pt}
\tablecaption{Host Spectroscopic Redshifts Without Available NED Information \label{tab:redshift}}
\tablehead{
\colhead{SN\tablenotemark{a}} & 
\colhead{z$_{\rm SN}$}   & 
\colhead{Host Galaxy} &
\colhead{z$_{\rm host}$}   &
\colhead{Telescope/Instrument} 
}
\startdata
ASASSN-15hx & N/A    & uncataloged &0.00812 & Magellan/IMACS \\
ASASSN-15rq & 0.025 &MRSS 349-038278& 0.02307 & Magellan/LDSS3 \\
ASASSN-15rw & 0.02  &WISEA J021558.50+121414.4& 0.01884 & F18\tablenotemark{b} \\
ASASSN-15ti & 0.016 & WISEA J030510.59+375359.9 & 0.01732 & F18\tablenotemark{b} \\
ASASSN-15uh & 0.0135 & KUG 0925+693 & 0.01489 & LBT/MODS \\
2016ag & 0.0187 &  WISEA J013123.32+601912.5& 0.01461 & Shane/KAST \\
2016asf & 0.021& KUG 0647+311 & 0.018019 & F18\tablenotemark{b} \\
2016blc & 0.012& WISEA J104848.74-201547.4 & 0.01285 & F18\tablenotemark{b} \\
\enddata
\tablenotetext{a}{The SN name adopts the IAU name when available or otherwise the survey name. All the IAU and survey names are available in Table~\ref{tab:sample}.}
\tablenotetext{b}{Host redshifts are obtained from the Foundation Supernova Survey \citep{Foley2018}}
\tablenotetext{c}{Host redshift of 2018fhw was first reported in the Astronomer's Telegram 11980 \citep{Eweis2018}}
\end{deluxetable}

\begin{deluxetable}{cccccccr}
\tabletypesize{\footnotesize}
\tablewidth{0pt}
\tablecaption{Optical Photometry Results\label{tab:lightcurves}}
\tablehead{
\colhead{SN}&
\colhead{JD} & 
\colhead{Mag}   & 
\colhead{Mag\_err}   &
\colhead{Filter}   &
\colhead{Sub$\tablenotemark{a}$}   &
\colhead{Isophote$\tablenotemark{b}$}   &
\colhead{Source} 
}
\startdata
2018jky & 2458456.7064 & 18.14 & 0.21 & $g$ & Y & N & ASAS-SN \\
2018jky & 2458457.6502 & 17.78 & 0.11 & $g$ & Y & N & ASAS-SN \\
2018jky & 2458458.7521 & 17.21 & 0.07 & $g$ & Y & N & ASAS-SN \\
2018jky & 2458459.7366 & 16.72 & 0.06 & $g$ & Y & N & ASAS-SN \\
2018jky & 2458461.6346 & 16.26 & 0.05 & $g$ & Y & N & ASAS-SN \\
2018jky & 2458462.7428 & 16.03 & 0.04 & $g$ & Y & N & ASAS-SN \\
2018jky & 2458463.4105 & 15.99 & 0.04 & $g$ & Y & N & ASAS-SN \\
2018jky & 2458459.7223 & 16.90 & 0.04 & $B$ & N & N & LCOGT1m \\
\enddata
\tablenotetext{a}{Whether image subtraction is used for photometry.}
\tablenotetext{b}{Whether the isophote model is used to subtract host-galaxy flux. See Section~\ref{sec:2ndstep} for detailed description on how the isophote model works for the photometry.}
\tablenotetext{}{\textbf{Notes.}  The entire table in machine-readable format is available in the online journal, and a portion is shown here for guidance regarding its form and content.}
\end{deluxetable}

\begin{longrotatetable}
\begin{deluxetable}{lcccccccccccc}
\setlength{\tabcolsep}{0.02in} 
\tabletypesize{\footnotesize}
\tablecolumns{13}
\tablewidth{0pt}
\tablecaption{Light-Curve Parameters from SNooPy Template Fitting \label{tab:lcparam_snpy_model}}
\tablehead{
\colhead{}& 
\colhead{}& 
\colhead{}&
\colhead{}& 
\colhead{{\normalsize\it{max\_model}}}&
\colhead{}& 
\colhead{}&
\colhead{}& 
\colhead{}&
\colhead{{\normalsize\it{EBV\_model2}}}&
\colhead{}&
\colhead{}&
\colhead{}\\
\cmidrule(r){2-8} 
\cmidrule(r){9-12} 
\colhead{SN\tablenotemark{a}} & 
\colhead{$t_{\rm peak}(B)$}   & 
\colhead{$s_{BV}$} &
\colhead{$B_{\rm peak}$}   &
\colhead{$g_{\rm peak}$} & 
\colhead{$V_{\rm peak}$}   & 
\colhead{$r_{\rm peak}$}   &
\colhead{$i_{\rm peak}$} &
\colhead{$t_{\rm peak}(B)$}&
\colhead{$s_{BV}$}&
\colhead{E$(B-V)_{\rm host}$}&
\colhead{$\mu$} &
\colhead{Category$^*$} \\
\colhead{}&
\colhead{$-$2,457,000}&
\colhead{}&
\colhead{(mag)}&
\colhead{(mag)} &
\colhead{(mag)}&
\colhead{(mag)} &
\colhead{(mag)} &
\colhead{$-$2,457,000} &
\colhead{}&
\colhead{(mag)}&
\colhead{(mag)} &
\colhead{}
}
\startdata
ASASSN-15aj & 36.22$\pm$0.25 & 0.75$\pm$0.04 & 15.01$\pm$0.04 & ... & 14.77$\pm$0.02 & 14.77$\pm$0.03 & 15.15$\pm$0.03 & 36.29$\pm$0.22 & 0.74$\pm$0.04 & 0.13$\pm$0.04 & 33.22$\pm$0.05 & C2 \\
ASASSN-15ak & 38.15$\pm$0.31 & 0.95$\pm$0.06 & 14.80$\pm$0.08 & ... & 14.70$\pm$0.01 & 14.76$\pm$0.05 & 15.29$\pm$0.09 & 38.02$\pm$0.14 & 0.95$\pm$0.03 & 0.09$\pm$0.01 & 33.64$\pm$0.04 & C2 \\
ASASSN-15db & 76.37$\pm$0.30 & 0.86$\pm$0.05 & ... & ... & 14.55$\pm$0.03 & ... & ... & ... & ... & ... & ... & C1 \\
ASASSN-15eb & 82.57$\pm$0.74 & 1.02$\pm$0.09 & 15.89$\pm$0.06 & ... & 15.82$\pm$0.05 & 15.79$\pm$0.04 & 16.35$\pm$0.08 & 82.38$\pm$0.75 & 1.02$\pm$0.07 & -0.04$\pm$0.05 & 34.65$\pm$0.10 & C2 \\
2015F & 107.60$\pm$0.21 & 0.86$\pm$0.02 & ... & ... & 13.31$\pm$0.02 & ... & ... & ... & ... & ... & ... & C1 \\
2015bp & 113.10$\pm$0.13 & 0.71$\pm$0.02 & ... & ... & 13.90$\pm$0.01 & ... & ... & ... & ... & ... & ... & C0 \\
ASASSN-15ga & 116.66$\pm$0.15 & 0.45$\pm$0.02 & 15.77$\pm$0.05 & ... & 15.08$\pm$0.03 & 15.10$\pm$0.02 & 15.28$\pm$0.03 & 116.59$\pm$0.18 & 0.46$\pm$0.02 & 0.14$\pm$0.06 & 32.95$\pm$0.05 & C0 \\
ASASSN-15go & 125.14$\pm$0.53 & 0.60$\pm$0.21 & 16.42$\pm$0.13 & ... & 16.04$\pm$0.09 & 15.91$\pm$0.44 & 16.38$\pm$0.21 & 125.84$\pm$0.69 & 0.71$\pm$0.15 & 0.12$\pm$0.12 & 34.27$\pm$0.24 & C2 \\
\enddata
\end{deluxetable}
\tablenotetext{a}{The SN name adopts the IAU name when available or otherwise the survey name. All the IAU and survey names are available in Table~\ref{tab:sample}.}
\tablenotetext{\dagger}{These names are used for brevity, the same as in Table \ref{tab:sample}.}
\tablenotetext{*}{As explained below, the supernovae are categorized into different groups according to their data coverage in terms of both wavelength and phase, and how well they are fitted by the template light curves:}
\tablenotetext{}{
\begin{itemize}
\setlength\itemsep{-1em}
\item[C0:] Good data coverage and fitting results.\\
\item[C1:] Only ASAS-SN $V$-band data are available. SNooPy {\it EBV\_model2} fitting is not feasible due to lack of multi-color coverage to derive host extinction. {\it max\_model} fitting has been performed, yielding reasonable results of the peak times and magnitudes though the derived$s_{BV}$ should be used with caution.\\
\item[C2:] Only a small number ($\lesssim3$) of epochs with available multi-band data. Some of the supernovae only have multi-band light curves close to the $B$-band peak, which makes it challenging to derive the decline rate (or the light-curve width). Similarly with C1, {\it max\_model} fitting gives credible results of peak times and magnitudes while the $s_{BV}$ parameters should be used with caution. {\it EBV\_model2} fitting is performed for these targets, but the fitting results have large uncertainties owing to inadequate coverages.\\
\item[C3:] The first available multi-band data point obtained $\gtrsim30$\,d after the estimated $B$-band peak. The light-curve parameters from both {\it max\_model} and {\it EBV\_model2} should be used with caution.\\
\item[C4:] Decent light-curve coverage but not well-fitted by SNooPy templates. Therefore, the best-fit parameters, especially those from {\it EBV\_model2}, should not be trusted.
\end{itemize}
}
\end{longrotatetable}

\begin{longrotatetable}
\begin{deluxetable}{lcccccc}
\tabletypesize{\footnotesize}
\tablecolumns{7}
\tablewidth{0pt}
\tablecaption{Light-Curve Parameters From Gaussian Process Fitting\label{tab:lcparam}}
\tablehead{
\colhead{SN\tablenotemark{a}} & 
\colhead{$t_{\rm max}(B)$}   & 
\colhead{$B_{\rm peak}$}   &
\colhead{$\Delta m_{15}(B)$} & 
\colhead{$t_{\rm max}(V)$}   & 
\colhead{$V_{\rm peak}$}   &
\colhead{$\Delta m_{15}(V)$} \\
\colhead{}&
\colhead{$-$2,457,000}&
\colhead{(mag)}&
\colhead{(mag)} &
\colhead{$-$2,457,000}&
\colhead{(mag)} &
\colhead{(mag)}
}
\startdata
ASASSN-15aj & ... & ... & ... & 37.9$\pm$0.5 & 14.76$\pm$0.02 & 0.93$\pm$0.07 \\
ASASSN-15db & ... & ... & ... & 77.0$\pm$0.3 & 14.51$\pm$0.02 & 0.83$\pm$0.06 \\
2015F & ... & ... & ... & 108.4$\pm$0.6 & 13.33$\pm$0.02 & 0.74$\pm$0.04 \\
2015bp & ... & ... & ... & 112.7$\pm$0.3 & 13.90$\pm$0.01 & 0.82$\pm$0.03 \\
ASASSN-15hf & ... & ... & ... & 138.9$\pm$0.5 & 14.26$\pm$0.02 & 0.68$\pm$0.05 \\
ASASSN-15hx & 151.5$\pm$0.2 & 13.29$\pm$0.01 & ... & 153.1$\pm$0.5 & 13.37$\pm$0.01 & 0.63$\pm$0.04 \\
ASASSN-15jo & 169.0$\pm$0.7 & 15.69$\pm$0.02 & 1.87$\pm$0.09 & ... & ... & ... \\
ASASSN-15kp & 188.5$\pm$1.2 & 15.52$\pm$0.03 & 0.90$\pm$0.13 & 188.9$\pm$1.7 & 15.53$\pm$0.05 & 0.51$\pm$0.09 \\
ASASSN-15pl & ... & ... & ... & 290.5$\pm$0.9 & 15.16$\pm$0.03 & 0.73$\pm$0.06 \\
ASASSN-15pz & 307.2$\pm$0.6 & 14.24$\pm$0.02 & 0.67$\pm$0.06 & 307.2$\pm$0.6 & 14.26$\pm$0.01 & 0.39$\pm$0.04 \\
ASASSN-15qc & 299.4$\pm$0.7 & 15.86$\pm$0.02 & 1.01$\pm$0.08 & 303.4$\pm$0.5 & 15.60$\pm$0.01 & 0.72$\pm$0.03 \\
2015ao & 307.4$\pm$0.6 & 17.52$\pm$0.03 & 1.84$\pm$0.16 & 310.2$\pm$0.2 & 16.87$\pm$0.02 & 1.49$\pm$0.06 \\
ASASSN-15rq & 324.9$\pm$0.3 & 15.45$\pm$0.01 & 0.93$\pm$0.05 & 326.1$\pm$0.4 & 15.46$\pm$0.01 & 0.61$\pm$0.04 \\
ASASSN-15rw & 330.7$\pm$0.8 & 15.65$\pm$0.05 & 1.00$\pm$0.14 & 332.0$\pm$0.9 & 15.59$\pm$0.04 & 0.62$\pm$0.08 \\
2015ar & 351.1$\pm$1.2 & 15.45$\pm$0.04 & 1.25$\pm$0.16 & 354.3$\pm$0.6 & 15.28$\pm$0.03 & 0.93$\pm$0.07 \\
PSN J21505094-7020289 & 360.4$\pm$0.5 & 15.56$\pm$0.03 & 1.39$\pm$0.07 & 362.1$\pm$0.4 & 15.17$\pm$0.02 & 0.81$\pm$0.03 \\
ASASSN-15ti & 365.2$\pm$0.8 & 16.26$\pm$0.06 & 1.53$\pm$0.12 & ... & ... & ... \\
2015bd & ... & ... & ... & 348.5$\pm$0.7 & 15.19$\pm$0.02 & 0.65$\pm$0.05 \\
ASASSN-15uh & 387.0$\pm$1.4 & 15.57$\pm$0.07 & 0.85$\pm$0.19 & 389.3$\pm$1.4 & 15.24$\pm$0.06 & 0.57$\pm$0.16 \\
ASASSN-15ut & ... & ... & ... & 392.4$\pm$0.5 & 16.32$\pm$0.03 & 1.03$\pm$0.17 \\
2016bfu & 471.0$\pm$0.4 & 16.21$\pm$0.03 & 1.91$\pm$0.09 & 473.1$\pm$0.4 & 15.63$\pm$0.03 & 1.24$\pm$0.04 \\
2016blc & 490.1$\pm$0.9 & 14.71$\pm$0.03 & 0.96$\pm$0.10 & 491.0$\pm$0.8 & 14.74$\pm$0.03 & 0.64$\pm$0.06 \\
2016bln & ... & ... & ... & 502.0$\pm$0.8 & 16.01$\pm$0.04 & 0.67$\pm$0.08 \\
2016cbx & 513.4$\pm$1.3 & 16.69$\pm$0.06 & 0.92$\pm$0.19 & 516.1$\pm$1.0 & 16.52$\pm$0.06 & 0.74$\pm$0.10 \\
2016ccz & ... & ... & ... & 540.5$\pm$0.5 & 15.44$\pm$0.03 & 0.79$\pm$0.07 \\
2016coj & ... & ... & ... & 549.7$\pm$0.3 & 13.01$\pm$0.01 & 0.70$\pm$0.02 \\
2016daj & 593.1$\pm$1.0 & 16.58$\pm$0.07 & 1.12$\pm$0.14 & 593.6$\pm$1.6 & 16.57$\pm$0.05 & 0.67$\pm$0.13 \\
2016ekg & 610.3$\pm$0.3 & 15.03$\pm$0.03 & 1.12$\pm$0.07 & 611.3$\pm$0.5 & 15.07$\pm$0.02 & 0.66$\pm$0.05 \\
2016ekt & 602.0$\pm$1.2 & 14.77$\pm$0.02 & 0.81$\pm$0.11 & 605.0$\pm$0.2 & 14.72$\pm$0.01 & 0.66$\pm$0.02 \\
2016euj & 619.7$\pm$0.3 & 15.35$\pm$0.02 & 1.39$\pm$0.05 & 620.5$\pm$0.3 & 15.33$\pm$0.01 & 0.82$\pm$0.03 \\
2016fej & 636.1$\pm$0.4 & 13.90$\pm$0.02 & 0.93$\pm$0.06 & 638.0$\pm$0.5 & 13.93$\pm$0.02 & 0.62$\pm$0.04 \\
2016fff & 630.3$\pm$0.3 & 15.03$\pm$0.02 & 1.77$\pm$0.06 & 632.1$\pm$0.3 & 14.90$\pm$0.02 & 0.98$\pm$0.05 \\
2016fob & ... & ... & ... & 633.5$\pm$1.3 & 16.15$\pm$0.02 & 0.62$\pm$0.07 \\
2016gfk & ... & ... & ... & 646.9$\pm$0.9 & 16.77$\pm$0.03 & 0.84$\pm$0.08 \\
2016gsb & 672.1$\pm$0.5 & 14.52$\pm$0.03 & 1.13$\pm$0.08 & 674.4$\pm$0.6 & 14.45$\pm$0.03 & 0.66$\pm$0.06 \\
2016gsn & 672.3$\pm$0.5 & 15.25$\pm$0.02 & 1.09$\pm$0.06 & 672.9$\pm$0.3 & 15.08$\pm$0.01 & 0.68$\pm$0.03 \\
2016gtr & 667.1$\pm$1.4 & 15.61$\pm$0.02 & 0.89$\pm$0.12 & 670.8$\pm$0.6 & 15.58$\pm$0.02 & 0.69$\pm$0.06 \\
2016gxp & 685.7$\pm$0.6 & 15.10$\pm$0.02 & 1.07$\pm$0.07 & 688.5$\pm$0.8 & 14.87$\pm$0.02 & 0.61$\pm$0.06 \\
2016hli & 696.3$\pm$0.5 & 17.47$\pm$0.03 & 1.58$\pm$0.08 & 698.0$\pm$0.7 & 16.82$\pm$0.03 & 0.87$\pm$0.06 \\
2016hpw & 703.4$\pm$0.3 & 16.02$\pm$0.02 & 1.04$\pm$0.04 & 705.3$\pm$0.3 & 15.97$\pm$0.01 & 0.70$\pm$0.03 \\
2016hvl & 710.9$\pm$0.6 & 15.95$\pm$0.04 & 1.17$\pm$0.07 & 714.0$\pm$0.9 & 15.49$\pm$0.03 & 0.67$\pm$0.05 \\
2016huh & ... & ... & ... & 700.4$\pm$1.0 & 16.86$\pm$0.03 & 0.78$\pm$0.09 \\
2016igr & 726.9$\pm$0.4 & 15.31$\pm$0.02 & 1.16$\pm$0.05 & 728.1$\pm$0.5 & 15.32$\pm$0.02 & 0.69$\pm$0.04 \\
2016ins & 724.1$\pm$0.8 & 16.89$\pm$0.03 & 1.50$\pm$0.23 & 725.4$\pm$0.9 & 16.58$\pm$0.02 & 0.89$\pm$0.09 \\
2016ipf & 728.0$\pm$0.5 & 16.76$\pm$0.02 & 1.35$\pm$0.08 & 727.9$\pm$1.2 & 16.74$\pm$0.03 & 0.65$\pm$0.08 \\
2016jab & 749.5$\pm$0.8 & 16.06$\pm$0.02 & 0.93$\pm$0.08 & 751.5$\pm$0.5 & 15.97$\pm$0.02 & 0.64$\pm$0.04 \\
2017jl & 784.9$\pm$0.3 & 15.01$\pm$0.02 & 0.99$\pm$0.06 & 786.6$\pm$0.4 & 14.94$\pm$0.02 & 0.63$\pm$0.06 \\
2017yv & 795.7$\pm$0.6 & 15.74$\pm$0.04 & 1.15$\pm$0.09 & 797.2$\pm$0.6 & 15.59$\pm$0.04 & 0.72$\pm$0.07 \\
2017awk & 808.4$\pm$0.5 & 16.03$\pm$0.02 & 1.22$\pm$0.06 & 810.5$\pm$0.6 & 15.79$\pm$0.02 & 0.73$\pm$0.05 \\
2017azw & 817.4$\pm$0.5 & 14.99$\pm$0.02 & 0.94$\pm$0.07 & 817.7$\pm$0.4 & 14.95$\pm$0.02 & 0.65$\pm$0.04 \\
2017bkc & ... & ... & ... & 813.3$\pm$1.0 & 16.66$\pm$0.02 & 0.57$\pm$0.08 \\
2017cav & ... & ... & ... & 820.2$\pm$1.6 & 16.32$\pm$0.02 & 0.53$\pm$0.09 \\
2017cbr & 834.7$\pm$0.4 & 15.96$\pm$0.02 & 1.31$\pm$0.07 & 836.1$\pm$0.6 & 15.78$\pm$0.02 & 0.77$\pm$0.06 \\
2017cbv & 840.8$\pm$0.4 & 11.73$\pm$0.02 & 0.97$\pm$0.05 & 842.4$\pm$0.6 & 11.68$\pm$0.02 & 0.59$\pm$0.04 \\
2017cfd & 844.1$\pm$0.2 & 14.89$\pm$0.02 & 1.18$\pm$0.05 & 845.1$\pm$0.3 & 14.77$\pm$0.03 & 0.74$\pm$0.05 \\
2017ckq & 851.6$\pm$0.4 & 14.34$\pm$0.01 & 1.16$\pm$0.05 & 852.9$\pm$0.4 & 14.35$\pm$0.01 & 0.72$\pm$0.03 \\
2017cjr & 846.9$\pm$0.3 & 15.04$\pm$0.02 & 1.22$\pm$0.04 & 848.6$\pm$0.4 & 15.08$\pm$0.02 & 0.71$\pm$0.04 \\
2017cts & 858.0$\pm$0.5 & 15.79$\pm$0.04 & 1.35$\pm$0.06 & 859.4$\pm$0.7 & 15.79$\pm$0.02 & 0.73$\pm$0.05 \\
2017cze & ... & ... & ... & 859.0$\pm$0.5 & 15.83$\pm$0.02 & 1.02$\pm$0.05 \\
2017cyy & 870.1$\pm$0.4 & 14.84$\pm$0.03 & 1.09$\pm$0.07 & 871.4$\pm$0.6 & 14.74$\pm$0.03 & 0.60$\pm$0.06 \\
2017dei & 868.7$\pm$0.7 & 16.69$\pm$0.03 & 1.71$\pm$0.12 & 871.3$\pm$0.6 & 16.43$\pm$0.02 & 1.01$\pm$0.07 \\
2017dit & 881.4$\pm$0.3 & 15.90$\pm$0.02 & 1.43$\pm$0.07 & 883.4$\pm$0.4 & 15.88$\pm$0.02 & 0.79$\pm$0.08 \\
2017dps & 882.4$\pm$0.3 & 14.81$\pm$0.03 & 1.64$\pm$0.05 & 884.1$\pm$0.5 & 14.81$\pm$0.03 & 0.84$\pm$0.05 \\
2017drh & 891.5$\pm$0.4 & 17.11$\pm$0.03 & 1.33$\pm$0.07 & 892.5$\pm$0.4 & 15.76$\pm$0.03 & 0.76$\pm$0.05 \\
2017egb & 906.5$\pm$0.3 & 15.72$\pm$0.03 & 1.62$\pm$0.07 & 907.8$\pm$0.4 & 15.68$\pm$0.01 & 0.84$\pm$0.04 \\
2017ejb & 911.0$\pm$0.1 & 15.83$\pm$0.02 & 2.04$\pm$0.04 & 913.2$\pm$0.2 & 15.38$\pm$0.01 & 1.30$\pm$0.03 \\
2017ejw & 912.5$\pm$0.3 & 15.51$\pm$0.03 & 1.45$\pm$0.06 & 913.7$\pm$0.3 & 15.55$\pm$0.03 & 0.86$\pm$0.05 \\
2017ekr & 914.5$\pm$1.6 & 16.15$\pm$0.07 & 1.43$\pm$0.40 & 915.9$\pm$1.7 & 15.99$\pm$0.06 & 0.95$\pm$0.34 \\
2017emq & 917.5$\pm$0.3 & 14.41$\pm$0.02 & 1.36$\pm$0.05 & 919.9$\pm$0.5 & 14.17$\pm$0.02 & 0.73$\pm$0.05 \\
2017enx & 917.4$\pm$1.4 & 14.05$\pm$0.05 & 1.67$\pm$0.19 & 920.1$\pm$1.5 & 13.83$\pm$0.05 & 0.93$\pm$0.18 \\
2017erv & 923.8$\pm$0.7 & 15.89$\pm$0.02 & 1.04$\pm$0.09 & 927.2$\pm$0.5 & 15.67$\pm$0.02 & 0.72$\pm$0.04 \\
2017erp & 934.7$\pm$1.7 & 13.69$\pm$0.04 & 1.03$\pm$0.21 & 936.7$\pm$1.6 & 13.52$\pm$0.03 & 0.65$\pm$0.10 \\
2017ezd & 942.1$\pm$0.4 & 15.79$\pm$0.02 & 1.35$\pm$0.07 & 942.5$\pm$0.4 & 15.73$\pm$0.02 & 0.78$\pm$0.04 \\
2017evn & 936.1$\pm$1.4 & 15.40$\pm$0.05 & 1.02$\pm$0.16 & 936.9$\pm$1.1 & 15.34$\pm$0.02 & 0.67$\pm$0.09 \\
2017exo & 937.4$\pm$0.8 & 16.70$\pm$0.03 & 0.96$\pm$0.09 & 938.8$\pm$0.6 & 16.24$\pm$0.03 & 0.69$\pm$0.04 \\
2017fbj & 944.6$\pm$0.5 & 16.22$\pm$0.03 & 1.18$\pm$0.09 & 947.6$\pm$0.6 & 15.95$\pm$0.04 & 0.71$\pm$0.08 \\
2017ffv & 956.1$\pm$1.2 & 15.49$\pm$0.07 & 1.13$\pm$0.14 & 957.7$\pm$1.1 & 15.22$\pm$0.04 & 0.73$\pm$0.08 \\
2017fgc & 958.8$\pm$0.2 & 13.76$\pm$0.01 & 1.02$\pm$0.03 & 963.1$\pm$0.5 & 13.55$\pm$0.01 & 0.72$\pm$0.02 \\
2017fzy & 978.7$\pm$0.9 & 16.32$\pm$0.02 & 1.84$\pm$0.14 & 983.2$\pm$1.4 & 15.97$\pm$0.02 & 1.08$\pm$0.11 \\
2017fzw & 987.8$\pm$0.2 & 14.21$\pm$0.02 & 1.71$\pm$0.03 & 990.1$\pm$0.4 & 13.85$\pm$0.02 & 0.88$\pm$0.05 \\
2017gah & 984.9$\pm$0.2 & 15.03$\pm$0.03 & 1.72$\pm$0.06 & 987.4$\pm$0.3 & 14.57$\pm$0.02 & 0.98$\pm$0.05 \\
2017gjn & 1004.6$\pm$0.3 & 15.14$\pm$0.01 & 1.05$\pm$0.03 & 1005.1$\pm$0.4 & 15.06$\pm$0.01 & 0.65$\pm$0.03 \\
2017glq & 1016.2$\pm$0.3 & 14.29$\pm$0.01 & 1.23$\pm$0.04 & 1016.8$\pm$0.4 & 14.28$\pm$0.01 & 0.68$\pm$0.04 \\
2017glx & 1009.8$\pm$0.6 & 14.69$\pm$0.03 & 0.89$\pm$0.07 & 1012.2$\pm$0.7 & 14.53$\pm$0.02 & 0.71$\pm$0.04 \\
2017grw & 1017.5$\pm$0.4 & 15.59$\pm$0.02 & 1.51$\pm$0.06 & 1019.0$\pm$0.4 & 15.51$\pm$0.02 & 0.80$\pm$0.04 \\
2017guh & 1022.0$\pm$0.6 & 15.20$\pm$0.03 & 1.36$\pm$0.12 & 1024.2$\pm$0.5 & 15.18$\pm$0.01 & 0.74$\pm$0.04 \\
2017haf & 1034.8$\pm$0.2 & 15.46$\pm$0.01 & 1.19$\pm$0.03 & 1035.8$\pm$0.3 & 15.35$\pm$0.01 & 0.75$\pm$0.02 \\
2017hgz & 1044.3$\pm$0.3 & 15.21$\pm$0.01 & 1.46$\pm$0.05 & 1046.1$\pm$0.4 & 15.18$\pm$0.02 & 0.77$\pm$0.04 \\
2017hjw & 1056.0$\pm$0.3 & 16.21$\pm$0.02 & 0.97$\pm$0.03 & 1057.0$\pm$0.3 & 15.90$\pm$0.01 & 0.63$\pm$0.03 \\
2017hjy & 1056.2$\pm$0.3 & 15.54$\pm$0.02 & 1.21$\pm$0.05 & 1057.7$\pm$0.4 & 15.45$\pm$0.01 & 0.66$\pm$0.03 \\
2017hle & 1049.3$\pm$0.5 & 17.93$\pm$0.01 & 1.70$\pm$0.07 & 1053.1$\pm$0.2 & 16.99$\pm$0.02 & 1.37$\pm$0.03 \\
2017hoq & 1061.2$\pm$1.3 & 16.03$\pm$0.04 & 0.89$\pm$0.12 & 1063.8$\pm$0.6 & 15.95$\pm$0.02 & 0.70$\pm$0.04 \\
2017hou & 1056.5$\pm$0.9 & 18.22$\pm$0.02 & 0.94$\pm$0.10 & 1057.5$\pm$0.5 & 17.47$\pm$0.02 & 0.64$\pm$0.06 \\
2017hpa & 1065.8$\pm$0.5 & 15.51$\pm$0.02 & 1.05$\pm$0.06 & 1068.7$\pm$0.5 & 15.39$\pm$0.01 & 0.72$\pm$0.03 \\
2017igf & 1085.0$\pm$0.2 & 14.87$\pm$0.02 & 1.87$\pm$0.05 & 1086.4$\pm$0.2 & 14.59$\pm$0.01 & 1.05$\pm$0.03 \\
2017iji & 1081.7$\pm$0.9 & 15.16$\pm$0.02 & 1.26$\pm$0.10 & 1084.0$\pm$0.8 & 15.02$\pm$0.02 & 0.74$\pm$0.06 \\
2017isj & 1092.9$\pm$1.3 & 15.81$\pm$0.03 & 0.99$\pm$0.16 & 1096.7$\pm$0.5 & 15.59$\pm$0.04 & 0.74$\pm$0.05 \\
2017iyb & 1117.4$\pm$0.6 & 14.91$\pm$0.01 & 1.42$\pm$0.08 & 1119.7$\pm$0.4 & 14.82$\pm$0.02 & 0.82$\pm$0.04 \\
2017iyw & ... & ... & ... & 1109.1$\pm$1.2 & 15.62$\pm$0.02 & 0.76$\pm$0.08 \\
2017jav & 1118.1$\pm$0.2 & 15.98$\pm$0.03 & 1.81$\pm$0.04 & 1119.3$\pm$0.3 & 15.80$\pm$0.01 & 0.96$\pm$0.03 \\
2017jdx & 1119.0$\pm$0.4 & 15.93$\pm$0.02 & 1.07$\pm$0.05 & 1120.1$\pm$0.4 & 15.86$\pm$0.01 & 0.60$\pm$0.03 \\
2018gl & 1138.8$\pm$0.1 & 16.31$\pm$0.00 & 1.83$\pm$0.02 & 1140.3$\pm$0.1 & 16.14$\pm$0.00 & 1.08$\pm$0.01 \\
2018gv & 1149.6$\pm$0.7 & 12.91$\pm$0.03 & 0.84$\pm$0.08 & 1151.1$\pm$0.6 & 12.89$\pm$0.02 & 0.62$\pm$0.05 \\
2018kp & 1159.3$\pm$0.3 & 16.62$\pm$0.02 & 1.22$\pm$0.05 & 1161.6$\pm$0.4 & 16.09$\pm$0.01 & 0.66$\pm$0.03 \\
2018pv & 1163.9$\pm$0.4 & 13.27$\pm$0.03 & 1.85$\pm$0.08 & 1166.8$\pm$0.2 & 12.68$\pm$0.03 & 1.18$\pm$0.06 \\
2018pc & 1165.3$\pm$0.4 & 15.42$\pm$0.02 & 1.41$\pm$0.08 & 1166.5$\pm$0.4 & 15.08$\pm$0.01 & 0.77$\pm$0.04 \\
2018oh & 1163.2$\pm$0.3 & 14.29$\pm$0.01 & 0.99$\pm$0.04 & 1163.8$\pm$0.3 & 14.29$\pm$0.01 & 0.65$\pm$0.03 \\
2018vw & 1177.9$\pm$0.5 & 15.37$\pm$0.03 & 0.92$\pm$0.07 & 1179.4$\pm$0.6 & 15.44$\pm$0.02 & 0.62$\pm$0.04 \\
2018xx & 1183.9$\pm$0.3 & 14.46$\pm$0.02 & 1.37$\pm$0.05 & 1185.1$\pm$0.4 & 14.43$\pm$0.02 & 0.79$\pm$0.06 \\
2018yu & 1194.7$\pm$0.4 & 13.98$\pm$0.02 & 1.03$\pm$0.06 & 1195.8$\pm$0.4 & 13.94$\pm$0.01 & 0.67$\pm$0.04 \\
2018zz & 1190.7$\pm$0.2 & 15.10$\pm$0.02 & 1.91$\pm$0.03 & 1192.4$\pm$0.2 & 14.96$\pm$0.01 & 1.03$\pm$0.03 \\
2018apo & 1223.6$\pm$0.7 & 15.50$\pm$0.03 & 0.93$\pm$0.09 & 1226.2$\pm$0.7 & 15.26$\pm$0.03 & 0.62$\pm$0.08 \\
2018aoz & 1222.6$\pm$0.3 & 12.84$\pm$0.03 & 1.33$\pm$0.07 & 1223.1$\pm$0.3 & 12.84$\pm$0.02 & 0.79$\pm$0.04 \\
2018aqi & 1223.1$\pm$0.3 & 15.81$\pm$0.03 & 1.48$\pm$0.06 & 1225.4$\pm$0.3 & 15.58$\pm$0.02 & 0.85$\pm$0.04 \\
2018aye & 1243.9$\pm$0.5 & 15.70$\pm$0.02 & 1.04$\pm$0.06 & 1244.7$\pm$0.4 & 15.67$\pm$0.02 & 0.71$\pm$0.03 \\
2018big & 1264.5$\pm$0.4 & 15.91$\pm$0.01 & 0.98$\pm$0.05 & 1266.1$\pm$0.5 & 15.80$\pm$0.01 & 0.60$\pm$0.04 \\
2018bta & ... & ... & ... & 1269.2$\pm$0.7 & 15.47$\pm$0.03 & 0.73$\pm$0.08 \\
2018cqw & 1300.7$\pm$0.4 & 14.56$\pm$0.01 & 0.99$\pm$0.05 & 1301.6$\pm$0.4 & 14.45$\pm$0.01 & 0.62$\pm$0.03 \\
2018ctv & 1294.3$\pm$0.6 & 17.42$\pm$0.03 & 2.00$\pm$0.07 & 1296.5$\pm$0.7 & 16.70$\pm$0.03 & 1.36$\pm$0.06 \\
2018cuh & 1305.5$\pm$0.4 & 15.02$\pm$0.03 & 1.20$\pm$0.08 & 1306.3$\pm$0.3 & 15.03$\pm$0.02 & 0.67$\pm$0.05 \\
2018dda & 1313.7$\pm$0.9 & 15.62$\pm$0.04 & 0.98$\pm$0.11 & 1315.6$\pm$0.8 & 15.29$\pm$0.03 & 0.67$\pm$0.06 \\
2018eay & 1328.5$\pm$0.5 & 17.48$\pm$0.02 & 0.94$\pm$0.07 & 1330.6$\pm$0.8 & 16.69$\pm$0.02 & 0.63$\pm$0.05 \\
2018ebk & 1330.2$\pm$0.4 & 16.88$\pm$0.03 & 1.26$\pm$0.05 & 1332.1$\pm$0.4 & 16.32$\pm$0.02 & 0.73$\pm$0.03 \\
2018enc & 1345.5$\pm$1.3 & 16.10$\pm$0.04 & 0.89$\pm$0.13 & 1346.5$\pm$1.3 & 15.94$\pm$0.03 & 0.66$\pm$0.09 \\
2018eov & 1342.4$\pm$0.9 & 15.65$\pm$0.03 & 1.04$\pm$0.12 & 1343.7$\pm$0.6 & 15.42$\pm$0.02 & 0.65$\pm$0.05 \\
2018eqq & ... & ... & ... & 1341.3$\pm$1.0 & 15.30$\pm$0.02 & 0.68$\pm$0.05 \\
2018feb & 1362.4$\pm$0.3 & 15.26$\pm$0.02 & 1.11$\pm$0.05 & 1363.5$\pm$0.4 & 15.24$\pm$0.01 & 0.60$\pm$0.03 \\
2018fhw & 1357.5$\pm$0.2 & 16.67$\pm$0.02 & 1.98$\pm$0.04 & 1360.3$\pm$0.3 & 16.37$\pm$0.02 & 1.22$\pm$0.04 \\
2018fop & ... & ... & ... & 1366.7$\pm$1.2 & 15.57$\pm$0.03 & 0.55$\pm$0.08 \\
2018fnq & 1374.6$\pm$0.6 & 15.77$\pm$0.03 & 1.06$\pm$0.07 & 1375.0$\pm$0.7 & 15.69$\pm$0.02 & 0.65$\pm$0.05 \\
2018fuk & 1377.1$\pm$0.4 & 15.87$\pm$0.01 & 1.20$\pm$0.06 & 1378.9$\pm$0.4 & 15.75$\pm$0.02 & 0.68$\pm$0.04 \\
2018ghb & 1381.0$\pm$0.9 & 14.81$\pm$0.05 & 1.59$\pm$0.12 & 1384.4$\pm$0.7 & 14.53$\pm$0.03 & 0.97$\pm$0.07 \\
2018hib & 1415.1$\pm$0.3 & 15.25$\pm$0.02 & 1.27$\pm$0.05 & 1416.2$\pm$0.4 & 15.22$\pm$0.01 & 0.74$\pm$0.03 \\
2018htt & 1438.6$\pm$0.2 & 13.91$\pm$0.03 & 1.59$\pm$0.04 & 1439.5$\pm$0.3 & 13.91$\pm$0.03 & 0.85$\pm$0.04 \\
2018hrt & ... & ... & ... & 1433.8$\pm$0.5 & 17.32$\pm$0.02 & 0.82$\pm$0.06 \\
2018hsa & 1432.3$\pm$0.4 & 15.78$\pm$0.02 & 1.14$\pm$0.07 & 1434.6$\pm$0.8 & 15.65$\pm$0.02 & 0.69$\pm$0.06 \\
2018ilu & 1449.7$\pm$1.0 & 15.42$\pm$0.02 & 0.93$\pm$0.09 & 1451.9$\pm$0.6 & 15.36$\pm$0.01 & 0.62$\pm$0.04 \\
2018isq & 1448.8$\pm$0.4 & 16.89$\pm$0.02 & 1.95$\pm$0.08 & 1451.8$\pm$0.2 & 16.00$\pm$0.01 & 1.30$\pm$0.04 \\
2018iuu & 1459.8$\pm$0.3 & 15.48$\pm$0.02 & 1.21$\pm$0.04 & 1462.0$\pm$0.7 & 15.46$\pm$0.02 & 0.74$\pm$0.04 \\
2018jaj & 1463.4$\pm$0.4 & 15.52$\pm$0.01 & 0.99$\pm$0.04 & 1464.6$\pm$0.4 & 15.41$\pm$0.01 & 0.70$\pm$0.03 \\
2018jeo & 1454.7$\pm$1.0 & 16.07$\pm$0.02 & 1.04$\pm$0.11 & 1456.5$\pm$0.5 & 15.94$\pm$0.01 & 0.61$\pm$0.05 \\
2018jky & 1469.0$\pm$0.2 & 15.39$\pm$0.03 & 1.69$\pm$0.06 & 1470.2$\pm$0.3 & 15.31$\pm$0.02 & 0.89$\pm$0.04 \\
2018jmo & ... & ... & ... & 1475.7$\pm$0.9 & 16.53$\pm$0.04 & 0.91$\pm$0.09 \\
2018jov & 1475.2$\pm$0.4 & 16.21$\pm$0.02 & 1.10$\pm$0.05 & 1476.3$\pm$0.5 & 16.06$\pm$0.01 & 0.64$\pm$0.04 \\
2018jwi & 1479.1$\pm$0.3 & 15.04$\pm$0.02 & 1.10$\pm$0.04 & 1480.4$\pm$0.4 & 15.04$\pm$0.02 & 0.69$\pm$0.04 \\
2018kmu & 1480.9$\pm$1.4 & 16.42$\pm$0.06 & 0.88$\pm$0.18 & 1483.5$\pm$1.8 & 16.30$\pm$0.07 & 0.69$\pm$0.12 \\
2019np & 1509.6$\pm$0.5 & 13.42$\pm$0.02 & 1.00$\pm$0.05 & 1510.9$\pm$0.7 & 13.39$\pm$0.02 & 0.63$\pm$0.04 \\
2019so & 1507.2$\pm$0.2 & 17.21$\pm$0.02 & 1.96$\pm$0.04 & 1509.2$\pm$0.2 & 16.67$\pm$0.02 & 1.36$\pm$0.04 \\
PSN J140216.0-533228.8 & 1513.0$\pm$0.9 & 17.01$\pm$0.03 & 0.85$\pm$0.10 & 1515.2$\pm$0.8 & 16.26$\pm$0.02 & 0.59$\pm$0.06 \\
2019gbx & 1647.0$\pm$0.3 & 14.80$\pm$0.03 & 1.24$\pm$0.06 & 1647.8$\pm$0.4 & 14.78$\pm$0.02 & 0.80$\pm$0.04 \\
2019hxc & 1663.4$\pm$0.9 & 16.60$\pm$0.05 & 1.31$\pm$0.15 & 1665.7$\pm$1.5 & 16.44$\pm$0.04 & 0.73$\pm$0.08 \\
2019khf & 1680.3$\pm$0.4 & 15.19$\pm$0.03 & 1.12$\pm$0.06 & 1681.8$\pm$0.6 & 15.20$\pm$0.03 & 0.63$\pm$0.05 \\
2020ue & 1873.5$\pm$0.2 & 12.22$\pm$0.01 & 1.51$\pm$0.04 & 1874.1$\pm$0.3 & 12.16$\pm$0.01 & 0.86$\pm$0.03 \\
\enddata
\tablenotetext{a}{The SN name adopts the IAU name when available or otherwise the survey name. All the IAU and survey names are available in Table~\ref{tab:sample}.}
\end{deluxetable}
\end{longrotatetable}

\clearpage

\appendix

\section{Observing Protocol}
\label{protocol}

\begin{figure}[h]
\centering
\includegraphics[width=12cm]{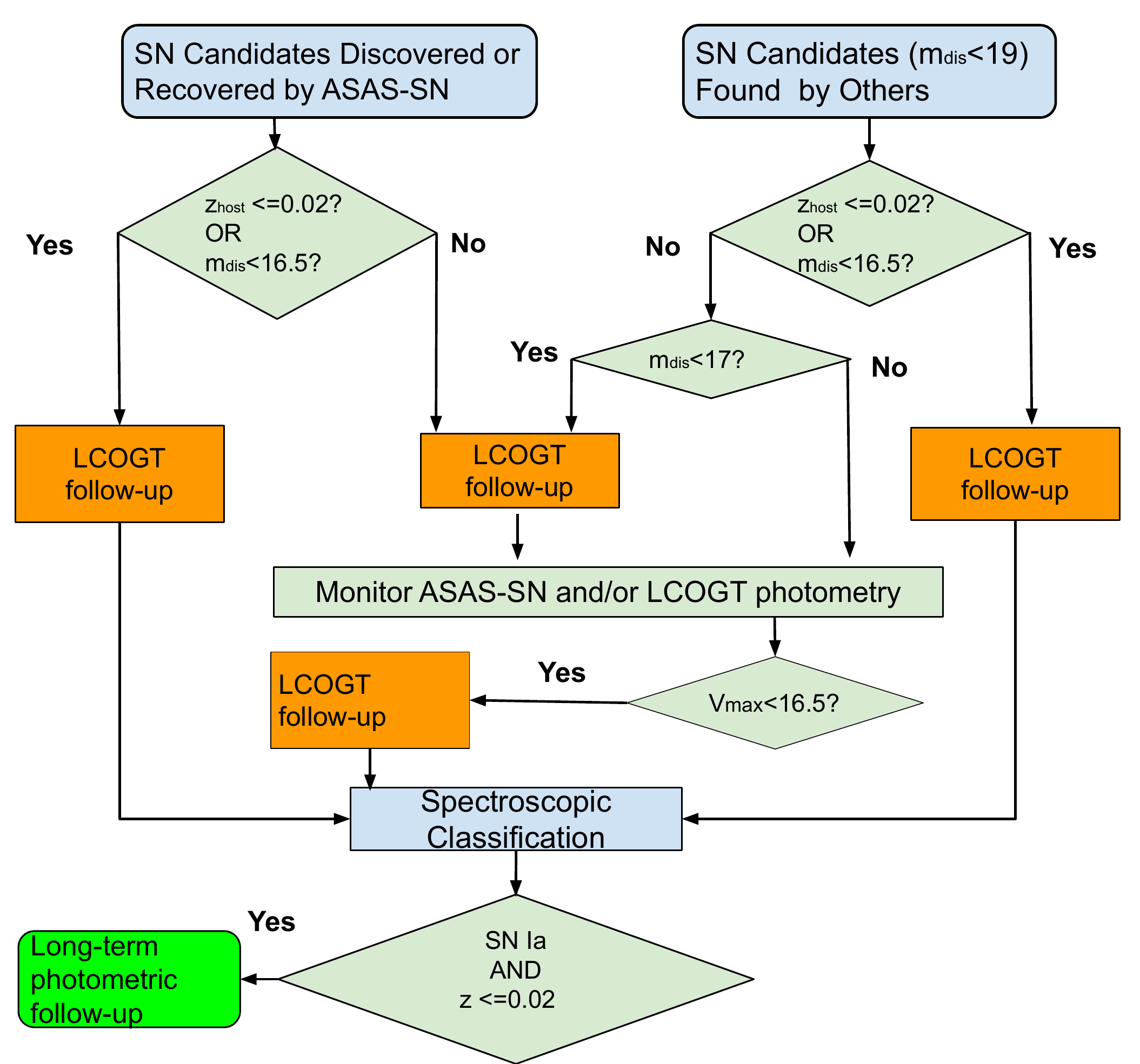}
\caption{Observing protocol for the CNIa0.02 complete sample.}
\label{fig:protocol}
\end{figure}

Fig.~\ref{fig:protocol} illustrates the protocol in our observing procedures when collecting our complete sample in the period between 2016 October and 2019 January. We scan the ASAS-SN transient page\footnote{\url{http://www.astronomy.ohio-state.edu/~assassin/transients.html}}, the Astronomer's Telegrams (ATel's\footnote{\url{http://www.astronomerstelegram.org/}}), the Transient Name Server (TNS\footnote{\url{https://wis-tns.weizmann.ac.il/}}), and the Bright Supernova webpage\footnote{\url{http://www.rochesterastronomy.org/snimages/}} on a daily basis, and record all bright transients with discovery magnitudes of $m_{\rm dis}<19$. To get early follow-up data for SNe~Ia in the sample, we start observations before classification for all potential targets according to the strategy in Figure~\ref{fig:protocol}. Meanwhile, we coordinate all available spectroscopic resources to classify the potential targets. Note that the primary aim of our complete sample is to include all spectroscopic sub-classes (e.g., 1991bg-like, 1991T-like) that belong to the SNe Ia population.  We made follow-up observations of some SNe Ia-like objects (e.g., SNe Iax) that are known to deviate from the WLR of SNe Ia, and they do not belong to the complete sample. For SNe without archival host redshifts, we followed up those with supernova redshifts of $z_{\rm SN} \lesssim 0.025$ if they have $V_{\rm peak}<16.5$. The selection of limit on $z_{\rm SN}$ is based on knowledge that the typical uncertainties of SN redshifts from spectroscopic classification are $\lesssim$ 0.005. SuperNova IDentification (SNID; \citealt{SNID}) is one of the commonly used tool for SNe classification. \cite{Stahl2020} investigated the SNID-determined redshifts by comparing them to the corresponding host-galaxy redshifts and found a standard deviation of 0.0039 for the difference between $z_{\rm SN}$ and $z_{\rm host}$.

\section{Follow-up Instruments}
\label{sec:instrument}
We used 1-m telescopes from the Las Cumbres Observatory Global Telescope network (LCOGT; \citealt{Brown2013}), which operates a number of robotic telescopes distributed at four sites (Siding Spring in Australia, Sutherland in South Africa, Cerro Tololo in Chile and McDonald Observatory in USA) covering both hemispheres. Each 1-m telescope is equipped with either a ``Sinistro'' or an SBIG STX-16803 camera\footnote{\url{https://lco.global/observatory/instruments/}}. The Bessel $BV$ and SDSS $ri$ filters are available on all telescopes, which were the main ones used for our optical observation. We also obtained some images using the 2-m or 0.4-m telescopes, and we plan to make those data available in the future. 

We used two 24-inch CDK24 telescopes operated by the Post Observatory (PO) mainly for following up northern objects. One is located at the Sierra Remote Observatories (SRO\footnote{\url{https://www.sierra-remote.com/}}; California, USA) and the other at Post Observatory Mayhill (New Mexico, USA). We used two types of cameras: an Apogee Alta U230 camera and an Apogee Alta U47 camera. Both cameras are back-illuminated, with similar quantum efficiency $>90\%$ over a broad region. The U230 camera was used by default at both sites. The telescope at SRO utilized the U230 for almost all images, and the Mayhill site had the U47 camera for a long period of time when U230 camera was unavailable for repairing its damaged shutter. Astrodon Photometrics $BV$\footnote{\url{https://astrodon.com/products/astrodon-photometrics-uvbri-filters/}} and Sloan $ri$\footnote{\url{https://astrodon.com/products/astrodon-photometrics-sloan-filters/}} filters were used.

\begin{table}
\caption{Instrument specifications}
\begin{center}
\begin{tabular}{lcccccc}
\hline
 Imager                  &Format& Binning      & Pixel scale           & Field of view                                   & Read Noise   & Gain\\
                              & (pixels)           & (pixels)                  &(arcsec/pixel)  & (arcmin $\times$ arcmin)                & (e$^-$)          & (e$^-$/ADU) \\
\hline
 SBIG STX-16803          &4096$\times$4096& 2$\times$2 & 0.464  & $15.8\times15.8$   & 13.5 & 1.5\\
 Sinistro                          &4096$\times$4096& 1$\times$1 & 0.389  & $26.5\times 26.5$ &7--8 & 1.0\\
Apogee Alta U230          &2048$\times$2048\tablenotemark{a}& 1$\times$1 & 0.77     & $13.1\times13.1$ &2.9&12.4\\           
Apogee Alta U47            &1024$\times$1024& 1$\times$1 & 0.67     & $11.4\times11.4$ &2.22&11.2\\
ANDICAM CCD       &2048$\times$2048&  2$\times$2 & 0.371   & $\sim 6 \times 6$ &6.5& 2.3\\
\hline
\end{tabular}
\label{tab:imager}
\end{center}
\tablenotetext{a}{The central 1024$\times$1024 pixels were used.}
\end{table}

The 1.3-m telescope of the Small \& Moderate Aperture Research Telescope System (SMARTS; \citealt{Subasavage2010}) is located at Cerro Tololo Inter-American Observatory (CTIO). It is equipped with A Novel Dual Imaging CAMera (ANDICAM; \citealt{DePoy2003}). The optical CCD for ANDICAM is a Fairchild 447 $2048 \times 2048$ pixel CCD. The IR Array for the ANDICAM is a Rockwell 1024x1024 HgCdTe ``Hawaii'' Array with 18-micron pixels. SMARTS/ANDICAM is equipped with standard KPNO-recipe Johnson-Kron-Cousins $BVRI$ filters and standard CIT/CTIO $JHK$ filters. SNe were observed in $BVRI$ and $JH$ bands with ANDICAM. In this data release, we publish $BV$ data. The $RI$ and $JH$ data taken by ANDICAM will be published in a future data release. 

\begin{figure}
\center
\includegraphics[width=12cm]{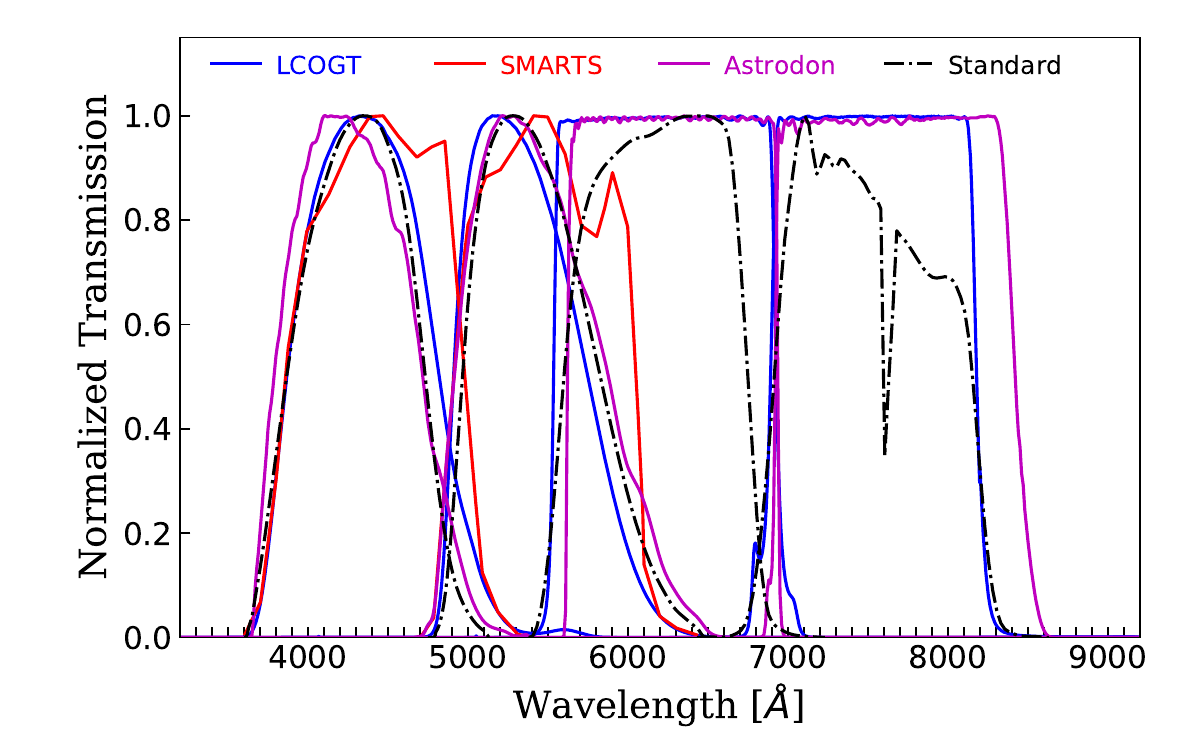}
\caption{The filter bandpasses used to obtain the majority of data released in this paper (solid lines). The standard Landolt $BV$ and SDSS $ri$ bandpasses (dashed lines) are shown for comparison.}
\label{fig:fset}
\end{figure}

The instruments described above are the primary ones used for DR1. Their instrument specifications are listed in Table~\ref{tab:imager}. The filter set used for observations in DR1 are compared to Landolt $BV$ \citep{Landolt1992} and the SDSS $ri$ \citep{Fukugita1996} standard bandpasses in Figure~\ref{fig:fset}.

DEdicated MONitor of EXotransits and Transients (DEMONEXT; \citealt{Villanueva2018}) is a 0.5-m PlaneWave CDK20 f/6.8 Corrected Dall-Kirkham Astrograph telescope at Winer Observatory in Sonoita, Arizona. DEMONEXT has a 2048$\times$2048 pixel FLI Proline CCD3041 camera, with a $30\arcmin \times 30\arcmin$ field of view (FOV) and a pixel scale of 0.9$\arcsec$/pixel. DEMONEXT has a full suite of Bessel $BVRI$ and SDSS $griz$ filters. $BVri$ data for four SNe (2016hli, 2016gou, 2016gxp and 2017isq) are included in DR1. We also include photometry for three SNe (2017ghu, 2017hle and 2018ast) obtained with Liverpool Telescope (LT) IO:O instrument in DR1. IO:O is the optical imaging component of the IO (Infrared-Optical) suite of instruments. It is equipped with a 4096$\times$4112 pixel e2V CCD 231-84, with a $10\arcmin \times10\arcmin$ FoV and a pixel scale of $\sim 0.3\arcsec$/pixel with $2\times2$ binning. The 1-m telescope at WeiHai Observatory of Shandong University (WHO; \citealt{Hu2014}) was used to obtain $BVri$ images for ASASSN-15uh. It has a back-illuminated PIXIS 2048B CCD camera at the Cassegrain focus, providing a $12\arcmin \times12\arcmin$ FoV and a pixel scale of 0.35$\arcsec$/pixel. The 0.41-m f/3.3 reflector telescope at A77 observatory (Dauban, 04150 Banon, France) was used to obtain some $BV$ images for 2015ar. The telescope is equipped with an ST8XME CCD, and its pixel scale is $1.4\arcsec$/pixel. 

We used instruments mounted on the du Pont 2.5-meter telescope, the 2.4-m Hiltner telescope and the 2.56-m Nordic Optical Telescope (NOT) to image a number of SNe.  We used two cameras mounted on du Pont. One is called ``CCD'', which is a direct imaging camera with a $2048\times2048$ pixel SITe2K CCD with plate scale of $0.259\arcsec$/pixel and field of view of $8.85\arcmin \times 8.85\arcmin$. The other is the WF4K CCD for the Wide Field Reimaging CCD Camera (WFCCD), which has a $4064\times4064$ pixel CCD with a plate scale of $0.484\arcsec$/pixel and field of view of $25\arcmin$ in diameter. For Hiltner, we used Ohio State Multi-Object Spectrograph (OSMOS), which is a wide field imager and multi-object spectrograph. For our imaging observation, a $4096\times4096$ pixel STA0500A CCD was used, which has a scale of $0.273\arcsec$/pixel and a field of view with $20\,\arcmin$ in diameter. On NOT, we used the Alhambra Faint Object Spectrograph and Camera (ALFOSC), which has both spectroscopic and imaging capabilities. The ALFOSC imaging was performed using a CCD231-42-g-F61 back-illuminated CCD with a field of view of $6.4\arcmin\times6.4\arcmin$ and a plate scale of 0.2138\,$\arcsec$/pixel in the imaging mode.

\section{Sample completeness}
\label{sec:completeness}

Our targets are selected according to the detections of the ASAS-SN survey, so the sample completeness depends on ASAS-SN's detection efficiency, that is, the fraction of the occurred supernovae which are detected by ASAS-SN. \cite{Holoien2019} compiled a sample of all supernovae detected by ASAS-SN between 2014 May 01 and 2017 December 31, and they found that the integral completeness (i.e., the cumulative detection efficiency) of this total sample is $95\pm3\,\%$ at $m_{\rm peak}=16.5$ by comparing with Euclidean predictions. That sample included supernovae found before the implementation of machine learning algorithm at the end of 2014, which substantially increased the detection efficiency (see Figure 4 of \citealt{Holoien2019}). During the collection of our complete sample (i.e., 2015 September 17 to 2019 January 31), ASAS-SN survey had several upgrades, including the operation of the Cassius unit in the summer of 2015 and the further expansions of three new units (Leavitt, Paczynski and Payne-Gaposchkin) in late 2017, which increased the limiting magnitude from $\sim17$ to $\sim18.5$. Therefore, there are good reasons to believe that, during our complete sample collection, the cumulative detection efficiency of ASAS-SN was at least 95\%, and probably greater, at $m_{\rm peak} = 16.5\,{\rm mag}$.

We also conduct external checks with the concurrent detections by ZTF to verify our selection criteria. ZTF conducts wide-field survey of the northern sky at a limiting magnitude of $\sim20.5$ \citep{Kulkarni2016}, and the Bright Transient Survey (BTS) project of ZTF aims to construct a magnitude-limited complete sample of transients with spectroscopic classifications down to m$<18.5$ \citep{Fremling2020, ZTFbright}.  The public start time of BTS survey was 2018 June 1, so there were an overlapping period of 8 months (from 2018 June 01 to 2019 January 31) with our complete sample. We first check whether all SNe Ia with peak magnitudes bright than 16.5 in BTS are included in our complete sample. Using {\it ZTF Bright Transient Survey Sample Explorer}\footnote{\url{https://sites.astro.caltech.edu/ztf/bts/explorer.php}}, we queried all transients with classification of ``SN Ia''  peaked between 2018 June 01 and 2019 January 31 with ${\rm peakmag}<16.5$ and $z\leq0.02$ and obtained a list of 14 objects. Among them, 2018dzy has a redshift $z_{\rm SN}=0.02$ based on ZTF SN classification spectrum, while its host galaxy (UGC 11873) is at $z=0.024760$ according to NED and was thus excluded from the CNIa0.02 complete sample. All others were included in CNIa0.02 and followed up by us. Our spectroscopic observations of the host galaxies show that four of them (2018fop, 2018htw, 2018jmo, 2018kmu) have host-galaxy redshifts $z_{\rm host}>0.02$. And all the other 9 SNe Ia (2018eay, 2018feb, 2018htt, 2018hkq, 2018iuu, 2018jaj, 2018jky, 2018jov, and 2019np) are included in the CNIa0.02 complete sample. In addition, we also queried all targets with classification as ``Candidate transients'' with $peakmag<16.5$ that were found between 2017 September 3 to 2019 December 31, and obtained 117 ZTF transients with peak magnitude brighter than 16.5. Among them, AT 2019ump (ZTF19acqnmjo) is likely a false positive (i.e., having a high probability as``bogus'' according to {\it AleRCE ZTF Explorer}\footnote{\url{https://alerce.online/object/ZTF19acqnmjo}}), and all the other 116 ZTF transient candidates were detected by ASAS-SN. Our cross examinations support that our sample is $\sim 100\%$ complete down to 16.5 mag.

{ \section{Exclusion of Peculiar Ia-like Objects}
\label{sec:purity}

The complete sample of CNIa0.02 include all spectroscopic subclasses of SNe Ia that are known to follow the width-luminosity relation (WLR) of SNe Ia (see \citealt{Phillips2017} for a recent review of WLR). These include the luminous Ia-91T subclass and low-luminosity Ia-91bg subclass that are sometimes reported as ``Ia-pec'' in classification reports due to historical reasons, but recent studies firmly show that they follow the WLR \citep{Phillips2017, Burns2018}, so they are  included in our complete sample. During the collection of our complete sample, we analyze the spectra using SNID \citep{SNID} to screen peculiar supernovae and also combine photometric properties when necessary to check whether a SN is truly peculiar according to our definition. Below we summarize peculiar supernovae that are excluded from our complete sample. 

ASASSN-15us was classified as SN Ia with the best-match to SN 2006bt, which is shown to be a peculiar SN Ia \citep{Foley2010}. Moreover, the late-time spectra of ASASSN-15us show significant peculiarities compared to normal SNe Ia, and we plan to present the detailed analysis of this SN in the future. 

ASASSN-15ut was first classified as SN Ia-91T \citep{Firth2016_as15ut}, but a later classification based on new spectroscopic observation reported that this is a peculiar object, challenging to tell whether it is a peculiar Ia or Type-Ib/c supernova \citep{Milisavljevic2016}. \cite{Holoien2017b} classified ASASSN-15ut as a Type-Ib/c supernova. We measure the peak time in $i$ and $B$ band for ASASSN-15ut and get the time of $i$-band primary maximum relative to the $B$-band $t^{i-B}_{max}\gtrsim8$\,d, which is outside the range for the SNe Ia population \citep{Ashall2020}.  

ASASSN-15pz is an overly-luminous supernova belonging to the ``2009dc-like SN Ia-pec'' group \citep{Chen2019}.  There is a large peak luminosity diversity within the group, and they generally do not follow the WLR. 

2016gxp was reported to match well with pre-maximum spectra of SN 2007if as well as young 91T-like SN Ia \citep{Reynolds2016}. It lacks a secondary maximum in the  $i$-band light curve and has $t^{i-B}_{max}=3.4\pm1.1$\,days, which do not favor the 91T-like SN Ia classification \citep{Ashall2020}. We plan to present the detailed analysis of 2016gxp in the future.

2017gbb is a Type Iax (02cx-like) supernova \citep{Lyman2017}, which does not belong to the SN Ia population.
}
\end{document}